\documentclass[sn-mathphys]{sn-jnl}

\usepackage[utf8]{inputenc}
\usepackage[]{hyperref}
\usepackage[]{graphicx}
\usepackage{amsmath,amssymb,xfrac,multirow}
\usepackage{float}
\usepackage{xcolor}
\usepackage{siunitx}
\usepackage{appendix}
\usepackage{miller}
\usepackage{scalerel}
\usepackage{verbatim}
\usepackage{cleveref}
\usepackage{soul}
\usepackage[T1]{fontenc}

\crefname{figure}{}{}
\Crefname{figure}{}{}
\newcommand{\etal}{\textit{et al}.}
\newcommand{\ie}{\textit{i.e.}}

\newcommand{\degree}{$^{\circ}$}
\newcommand{\wustite}{Fe$_{1-x}$O}
\newcommand{\wustiteSlab}{FeO}
\newcommand{\magnetite}{Fe$_3$O$_4$}
\newcommand{\hematite}{Fe$_2$O$_3$}

\newcommand{\RV}[1]{{\textcolor{black}{#1}}}

\usepackage{caption}
\usepackage{xparse}
\renewcommand{\thetable}{\arabic{table}} 
\renewcommand{\thefigure}{\arabic{figure}}
\newcommand{\Fig}[2]{Fig.~\ref{#1}\IfValueT{#2}{\MakeLowercase{#2}}}                    
\newcommand{\Figs}[2]{Figs.~\Cref{#1}\IfValueT{#2}{\MakeLowercase{#2}}}                 
\newcommand{\Figure}[2]{Figure~\ref{#1}\IfValueT{#2}{\MakeLowercase{#2}}}               
\newcommand{\Figures}[2]{Figures~\Cref{#1}\IfValueT{#2}{\MakeLowercase{#2}}}            
\newcommand{\fig}[2]{Supplementary Fig.~\ref{#1}\IfValueT{#2}{\MakeLowercase{#2}}}      
\newcommand{\figs}[2]{Supplementary Figs.~\Cref{#1}\IfValueT{#2}{\MakeLowercase{#2}}}   
\newcommand{\Tab}[1]{Supplementary Table~\ref{#1}}

\jyear{2024}%

\theoremstyle{thmstyleone}%
\theoremstyle{thmstyletwo}%
\theoremstyle{thmstylethree}%
\begin{document}

\title{Complexions at the Iron-Magnetite Interface}

\author*[1]{\fnm{Xuyang} \sur{Zhou}}\email{x.zhou@mpie.de}
\author*[1]{\fnm{Baptiste} \sur{Bienvenu}}\email{b.bienvenu@mpie.de}
\author[1]{\fnm{Yuxiang} \sur{Wu}}\email{yuxiang.wu@mpie.de}
\author[1]{\fnm{Alisson} \sur{Kwiatkowski da Silva}}\email{a.kwdasilva@mpie.de}
\author[2]{\fnm{Colin} \sur{Ophus}}\email{clophus@lbl.gov}
\author*[1]{\fnm{Dierk} \sur{Raabe}}\email{raabe@mpie.de}

\affil[1]{\orgname{Max-Planck-Institut for Sustainable Materials (Max-Planck-Institut für Eisenforschung)}, \orgaddress{\street{Max-Planck-Straße 1}, \city{Düsseldorf}, \postcode{40237}, \country{Germany}}}
\affil[2]{\orgname{National Center for Electron Microscopy, The Molecular Foundry, Lawrence Berkeley National Laboratory}, \orgaddress{\city{Berkeley}, \postcode{94720}, \country{USA}}}

\abstract{
Synthesizing distinct phases and controlling the crystalline defects in them are key concepts in materials and process design.
These approaches are usually described by decoupled theories, with the former resting on equilibrium thermodynamics and the latter on nonequilibrium kinetics.   
By combining them into a holistic form of defect phase diagrams, we can apply phase equilibrium models to the thermodynamic evaluation of defects such as vacancies, dislocations, surfaces, grain boundaries, and phase boundaries, placing the understanding of material imperfections and their role on properties on solid thermodynamic and theoretical grounds. 
In this study, we characterize an interface-stabilized phase between Fe and {\magnetite} (magnetite) with differential phase contrast (DPC) imaging in scanning transmission electron microscopy (STEM). 
This method uniquely enables the simultaneous imaging of both heavy Fe atoms and light O atoms, providing precise mapping of the atomic structure and chemical composition at this heterogeneous metal-oxide interface.
We identify a well-ordered two-layer interface-stabilized phase state (referred to as complexion) at the Fe\hkl[001]/{\magnetite}\hkl[001] interface. 
Using density-functional theory (DFT), we not only explain the observed complexion but also map out various interface-stabilized phases as a function of the O chemical potential. 
We show that the formation of complexions influences the properties of the interface, increasing its adhesion by 20~\% and changing the charge transfer between adjacent materials, also leveraging impact on the transport properties across such interfaces.
Our findings highlight the potential of tunable phase states at defects as a new asset in advanced materials design, paving the way for knowledge-based and optimized corrosion protection, catalysis, magnetism, and redox-driven phase transitions. 
These advancements can help to make the materials sector more sustainable, encompassing challenges such as longevity, efficient conversion, electrification and green steel production.
}


\maketitle

Iron oxides, ubiquitous in nature, science and engineering as well as integral to Earth's geological evolution, have far-reaching impacts on human civilization \cite{navrotsky2008Review,young2023Geo,tartaj2011Review}.
Their high importance includes topics such as heterogeneous catalysis \cite{weiss2002Review,parkinson2016Review}, corrosion \cite{toney1997Corrosion}, spintronics \cite{coey2003Review}, magnetic recording \cite{bate1991Review}, energy \cite{zhang2014Battery}, medicine \cite{laurent2008Review}, extraplanetary and terrestrial geophysics \cite{young2023Geo}, geology \cite{Dobson2005Geo,armstrong2019Geo}, mining \cite{devlin2023Mine} and the emerging field of sustainable metallurgy \cite{raabe2019Review,raabe2023Review,zhang2023HemMag,zhou2023Pore}.
The interfaces that connect iron oxides with adjacent phases play crucial roles for thermodynamics, kinetics, properties and performance across many types of practical applications \cite{mulakaluri2009Water,bliem2014SCV,parkin1986Dead,toney1997Corrosion,parkinson2016Review,raabe2023Review,kim2021HBD,pratt2012HSurface,hamed2019Magnetic,liu2021BiXMag,hosseini2018Redox,santos2014Redox,zhang2023HemMag,ma2022HBD}. 
Gaining deep understanding of the phenomena taking place at and across these interfaces is essential for fully exploiting their potential and controlling the thermodynamics and kinetics involved, leveraging the desired structure and chemistry for optimal performance, or respectively better understanding in each of these specific fields.

Deciphering the unexpected complexity of their atomistic and chemical structure makes such interfaces accessible to manipulation, for example with respect to their role and properties in chemical redox processes. 
The iron oxide surface, for example, undergoes reconstruction—a process predominantly governed by Tasker's criteria for surface polarity \cite{tasker1979Surface} and by the prevalence of oxygen vacancies \cite{bliem2014SCV}.
The atomic configurations of the surface terminations are not fixed, but they adapt based on the chemical potential of O in their immediate environment \cite{bliem2014SCV}. 
The interfaces between oxides and other phases are not merely kinematic entities with predetermined geometrical and chemical conditions. 
Instead, they can be subject to complex chemical, structural and electronic variations and reconstructions beyond singular atomistically sharp transition features, as revealed by recent observations \cite{bliem2014SCV,baram2011IF}.

One particularly interesting class of interface reconstructions is the so-called complexion \cite{dillon2007Concept,cantwell2014Review}. 
These have been described as thermodynamic entities \cite{korte2022Review} that form a compliant ``mediating'' layer with generic structural and chemical phase-like features between adjacent phases, different from the kinematically expected interface structure and chemistry. 
These complexions were suggested to not just represent a kinetic phenomenon but a thermodynamic one, that is only existent in the presence of adjacent phases, but not in conventional bulk phase diagrams, which are, per definition, defect-free representations of equilibrium phase states as a function of concentration, temperature, chemical potential and/or pressure.
Although ``complexion'' has gained acceptance in various applications \cite{baram2011IF,khalajhedayati2016Complexion,kuzmina2015Dislocation,meiners2020CuGB,yu2017NiBiGB}, its terminology still invites debate \cite{schusteritsch2021Dispute,NM2021Dispute}. 
 In this work, we consistently use ``complexion'' aligning with our experimental observations and recent reviews \cite{dillon2007Concept,cantwell2014Review} to describe these interfacial phenomena accurately.
The atomic structure and chemistry at these interfaces differ significantly from their bulk counterparts and can even undergo structural and chemical transitions, similar to bulk phase transformations \cite{buban2006Al2O3,baram2011IF,wang2011Order,luo2011LME,nie2013MgTwin,khalajhedayati2016CuZr,yu2017NiBiGB,meiners2020CuGB}. 
Given that the chemical potentials of all species i involved, denoted as $\mu_i$, remain constant across complexions adjacent to phases, it becomes feasible to construct complexion diagrams using $\mu_i$ as the primary variable, alongside appropriate thermodynamic parameters that govern the stability and transitions of such defect structures \cite{wang1998Hema,pentcheva2005JT,lodziana2007Verwey,mulakaluri2009Water,novotny2013FeRich,santos2014Redox,noh2015Mag111,ophus2013IF}. 
Such a complexion diagram becomes an invaluable tool for investigating defect phases and the transitions among them, bringing the fields of lattice defects and thermodynamics closely together.

The importance of structural and chemical features of these interfacial reconstructions, or defect phase states, has been shown for a number of systems. 
The Dillon-Harmer complexions, categorized into six distinct types and originally observed in undoped and doped Al$_2$O$_3$ ceramics, helped to elucidate the mechanisms underlying the longstanding question of abnormal grain growth in inorganic systems \cite{dillon2007Concept}. 
Rare-earth elements, such as La, preferentially segregate to the grain boundaries of Si$_3$N$_4$ to form nanometer-scale amorphous inter-granular films, essential in developing a microstructure with elongated grains that play a key role in tuning mechanical properties \cite{shibata2004Si3N4,ziegler2004Si3N4}. 
In addition to ceramics, metallic systems also exhibit comparable phenomena, such as the formation of a bi-layer interfacial phase of Bi at Ni grain boundaries, a characteristic associated with the susceptibility to liquid metal embrittlement \cite{luo2011LME}.
In the case of the Au/Al$_2$O$_3$ interface, nanometer-thick amorphous intergranular films have been shown to reach equilibrium at their interfaces \cite{baram2011IF}.
These examples illustrate the broad significance of complexions in materials science and beyond, potentially leveraging substantial influence on material properties under various types of boundary conditions.

While studying the complexities of interface structures yield significant insights, their detailed characterization—specifically for iron oxides—poses considerable challenges, potentially leading to their under-representation in earlier investigations.
Conventional scanning transmission electron microscopy (STEM) methods such as high-angle annular dark field (HAADF) imaging have limitations in resolving light elements, like O, against heavier ones, like Fe. 
In addressing this challenge, a recent breakthrough with the differential phase contrast (DPC) - four-dimensional STEM (4DSTEM) method has emerged, significantly enhancing the atomic-scale mapping of both structural and chemical features in such reconstructed regions, where imaging of these details is required to reveal the thermodynamic nature of interfacial phase states \cite{shibata2012DPC,muller2014DPC,muller2017DPC,hachtel2018DPC,gao2019DPC,ophus20194D}. 
This technique advances the study of complex interfaces at high spatial resolution, offering unprecedented insights into their structures and properties \cite{gao2019DPC}. 
Notably, the DPC-4DSTEM approach enables the reconstruction of charge-density maps from the collected data, providing phase contrast that correlates linearly with the mass of the elements. 
This enables direct spatial resolution of light atoms together with heavier atoms \cite{zheng2021H,zhou2023FeBC}, such as O within oxides and at interfaces. 
In the following sections, we present how DPC-4DSTEM probing advances our understanding of complex interface phenomena, offering insights that can lead to informed strategies for material design and a better understanding of geophysical phenomena, catalysis and sustainable metallurgical synthesis.

\subsubsection*{Mapping the structure of Fe\hkl[001]/{\magnetite}\hkl[001] interfaces at the atomic scale} \label{atomic_interfacial}

We conducted DPC-4DSTEM measurements to study the atomic structure of the Fe/iron oxide interface, formed by the epitaxial growth of multi-layer thin films on single-crystal MgO substrates at 300\,{\degree}C. 
We chose the thin film method to fabricate interfaces with a specific orientation relationship, aiming to create an ideal sample condition for high-resolution imaging. 
Further details on sample preparation via physical vapor deposition (PVD) can be found in Methods Section~\nameref{thin_film_synthesis} and \figs{fig:EDS,fig:4DSTEM,fig:STEM_IF}{}.

Using the {\magnetite} oriented in the \hkl[110] direction, we demonstrated the strength of the DPC-4DSTEM method in resolving both heavy Fe and light O atomic columns simultaneously.
\Figures{fig:fig1}{a}-i-iv showcase the corresponding experimentally reconstructed virtual dark-field image, electric field vector map, projected electrostatic potential map, and charge-density map from the DPC-4DSTEM data set.
The virtual dark-field image (see \Fig{fig:fig1}{a}-i) reveals the position of the Fe atomic columns, but it lacks sufficient resolution to distinguish lighter O atomic columns. 
On the other hand, the projected potential map (see \Fig{fig:fig1}{a}-iii) and the charge-density map (see \Fig{fig:fig1}{a}-iv) can clearly resolve both types of atomic columns.
In these three images, the positions of parts of the O atomic columns are indicated by white arrows.
Additional reconstructed images, including the center of mass, are presented in Methods Section~\nameref{Characterization} and \figs{fig:DPC_M13,fig:DPC_M18,fig:DPC_IM}{}.
Since the charge-density map provides the clearest simultaneous representation of both the heavy Fe and the lighter O atoms, we will primarily use it to investigate the atomic structure of the Fe/iron oxide interface in the subsequent discussion.

\begin{figure}[htbp!]
    \centering
    \includegraphics[trim = 1.0mm 0.5mm 2mm 2mm, clip, width=1.0\linewidth]{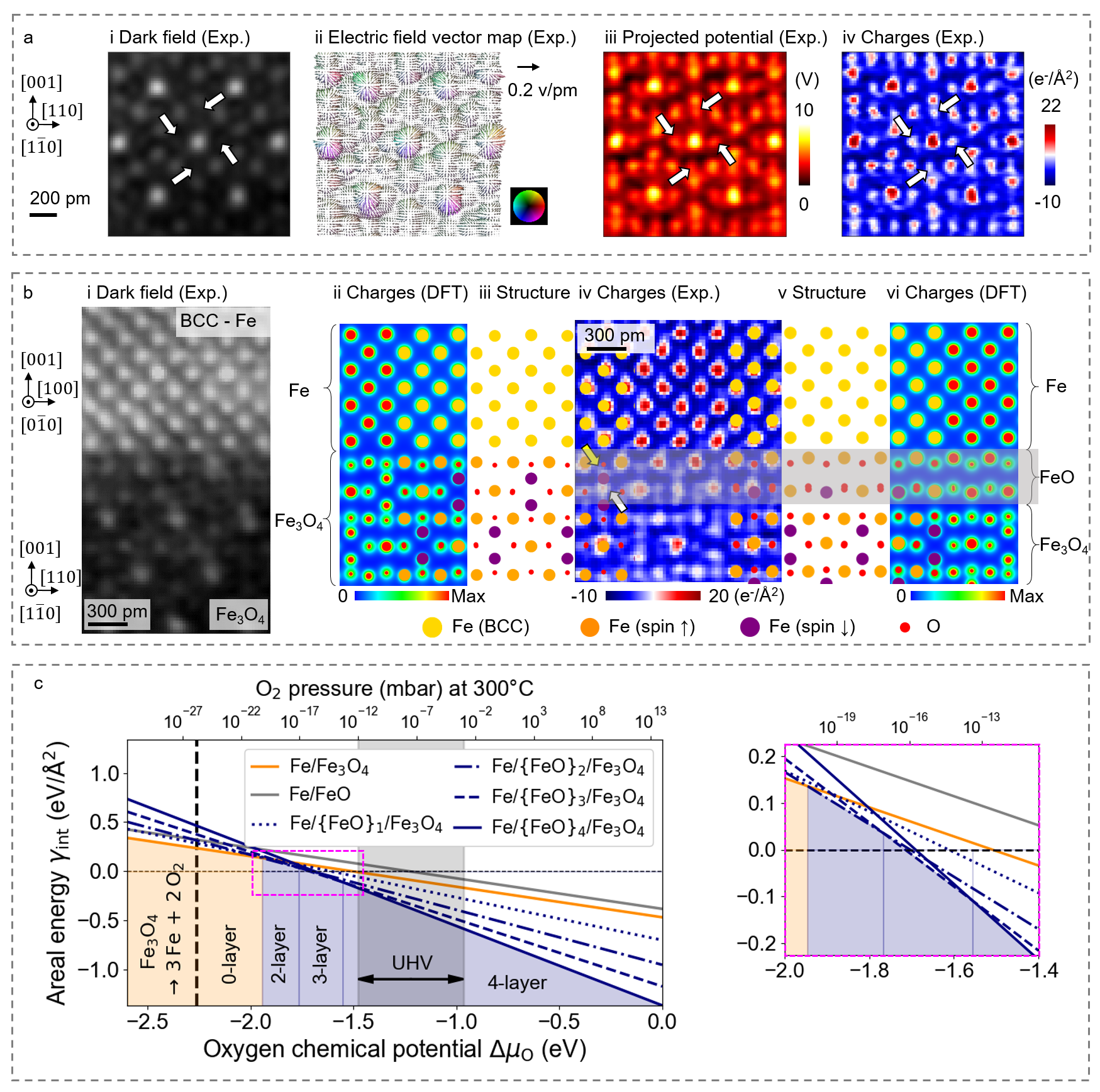}
    \caption{
    \textbf{The complexion diagram of the Fe/{\magnetite} interface.} 
    \textbf{a} Experimental (Exp.) differential phase contrast (DPC) - four-dimensional scanning transmission electron microscopy (4DSTEM) reconstruction for {\magnetite} oriented in the \hkl[110] direction: \textbf{i} Reconstructed virtual dark-field image \textbf{ii} Electric field vector map \textbf{iii} Projected electrostatic potential map \textbf{iv} Charge-density map. 
    The scanning step size used in this experiment is 13\,pm. 
    The white arrows in \textbf{i}, \textbf{iii}, \& \textbf{iv} indicate the positions of the O atomic columns. 
    \textbf{b} A combined experimental observation and theoretical study of the Fe\hkl[001]/{\magnetite}\hkl[001] interface includes: 
    experimental DPC-4DSTEM reconstructed \textbf{i} virtual dark-field image and \textbf{iv} charge-density map, along with results from DFT calculations comprising charge-density maps (\textbf{ii} \& \textbf{vi}) and relaxed structures (\textbf{iii} \& \textbf{v}) of the interface between Fe\hkl[001] and {\magnetite}\hkl[001]. 
    The first set of DFT calculations (\textbf{ii} \& \textbf{iii}) illustrates the pristine interface configuration, while the second set (\textbf{v} \& \textbf{vi}) depicts the interface with a reconstructed two-layer thick {\wustiteSlab} slab.
    In \textbf{iii} \& \textbf{v}, Fe atoms in the BCC structure are represented in yellow. 
    In the {\magnetite} structure, Fe atoms are color-coded based on their spin orientation: spin-up atoms are shown in purple and spin-
    }
    \label{fig:fig1}  
\end{figure}
\begin{figure}[htbp!]
    \ContinuedFloat 
    \caption*{down atoms in orange. 
    O atoms are depicted in red.
    In \textbf{iii}, yellow and white arrows highlight atomic columns that deviate from the positions observed experimentally in the {\magnetite} structure when the two-layer thick {\wustiteSlab} slab is not inserted.
    \textbf{c} DFT predicted complexion diagram versus the O chemical potential shift, with $\Delta \mu_{\text{O}}$ taking molecular $\text{O}_2$ as the reference upper limit, and dissociation of {\magnetite} into metallic Fe and molecular O$_2$ as the lower limit (vertical black dashed line).
    The explored interface configurations include: Fe/{\magnetite} (solid orange), Fe/{\wustiteSlab} (solid grey), and Fe/$\{ \text{\wustiteSlab} \}_n$/{\magnetite}, with $n$ being the number of {\wustiteSlab} layers (dark blue). 
    A horizontal dashed line is included in the graph as a visual reference.  
    The inset on the right shows a magnification in the range of chemical potentials where the transition occurs from 4 to 2 intermediate {\wustiteSlab} layers, separated by the vertical thin blue line.
    The O chemical potential is also converted to O$_2$ pressure at the 300\,{\degree}C temperature of the experiments, with the black shaded region showing typically accessible experimental conditions under ultra-high vacuum (UHV).
    }
\end{figure}

\Figures{fig:fig1}{b}-i \& iv show the reconstructed virtual dark-field image and charge-density map of the Fe\hkl[001]/{\magnetite}\hkl[001] interface, respectively.
In the charge-density map (see \Fig{fig:fig1}{b}-iv), the red and white pixels pinpoint atomic column locations, revealing an atomically sharp Fe\hkl[001]/{\magnetite}\hkl[001] interface. 
The upper part is body-centered cubic (BCC)-Fe.
The lower part consists of {\magnetite} (magnetite), which has an inverse spinel structure characterized by a face-centered cubic (FCC) sublattice of O\textsuperscript{2-} anions, with Fe\textsuperscript{2+} and Fe\textsuperscript{3+} cations occupying the interstitial sites \cite{verwey1939Magnetite}.
We observed no abrupt mono-layered transition plane between the two bulk phases; instead, we found a complex reconstruction layer, as characterized by the charge-density map in \Fig{fig:fig1}{b}-iv.
The reconstructed regions four atomic layers away from the interface can be clearly identified in both the upper and lower parts. 
However, accurately interpreting the atomic structure at the very Fe\hkl[001]/{\magnetite}\hkl[001] interface remains challenging when mapping these structural-chemical features only. 
Hence, in the next section, we derived the underlying complexion diagrams to better understand the reason for this reconstruction.

\subsubsection*{The Fe/{\magnetite} complexion and its associated complexion diagram} \label{complexion_DFT}

To better understand the structure of the reconstructed Fe\hkl[001]/{\magnetite}\hkl[001] interface region, we employed density functional theory (DFT) calculations to search for its thermodynamically most stable configuration. 
The results are presented in full detail in the \fig{fig:fig_sites_Fe_Fe3O4}{}. 
Considering the DFT structural parameters of BCC-Fe and {\magnetite} (see \Tab{tab:tab_DFT_bulk}), the stacking of the two layers results in an in-plane lattice mismatch of $\varepsilon_{xy}^{\text{BCC-Fe}}=-4.5\,\%$ and $\varepsilon_{xy}^{\text{\magnetite}}=+4.7\,\%$, with respect to the other layer, respectively.

Looking at the observed interface structure of \Figs{fig:fig1}{b}-i \& iv, the {\magnetite} layer appears to adopt a FeO$_2$\,-termination, and we thus mainly focused on this structure in the calculations.
We find the lowest energy configuration when BCC-Fe atoms are located at the alternating hollow sites of the upper FeO$_2$\,-terminated {\magnetite} layer. 
Here, ``hollow sites'' refer to positions that are typically in a concave arrangement, surrounded by other atoms, resulting in the strongest adhesion energy of approximately -3\,J/m$^2$, representing the energy required to separate the two slabs at their interface into two seperate ones with free surfaces.
The calculated charge-density and optimized structure for the most stable interface configuration generally align well with our experimental observations. 
This alignment is illustrated in the bridging sketch (see \Fig{fig:fig1}{b}-iii), which connects the experimental (see \Fig{fig:fig1}{b}-iv) and calculated (see \Fig{fig:fig1}{b}-ii) charge-density maps.
However, there are some discrepancies in certain areas.
Upon closer inspection of the experimental interface structure presented in \Fig{fig:fig1}{b}-iv, we noted a significant distortion on the {\magnetite} side, particularly within the first two atomic layers.
Firstly, in the first atomic layer (closest to the Fe), several O atomic columns (indicated by yellow arrows) appear to have shifted substantially towards the {\magnetite} side. 
A slight shift of these atomic columns, in the range of a few picometers, is also corroborated by our DFT calculations, as seen in \Fig{fig:fig1}{b}-ii, but not to a sufficient extent to explain this discrepancy.
Secondly, in the second atomic layer (the second one closest to the BCC-Fe), the Fe atoms located at the tetrahedral sites of the FCC-O sublattice (in purple on \Figs{fig:fig1}{b}-iv) have noticeably moved away from the Fe side.
Simultaneously, the Fe atoms in the fourth atomic layer of the {\magnetite} have shifted towards the Fe side. 
These two columns of Fe atoms appear to have merged into a single atomic column, as indicated by the white arrows in \Fig{fig:fig1}{b}-iv.

We propose that the region near the interface adopts a well-ordered two-layer interface-stabilized phase state (referred to as complexion), reminiscent of the iron oxide polymorph {\wustiteSlab} (wüstite). 
{\wustiteSlab} exhibits a NaCl structure and becomes thermodynamically stable above 570\,{\degree}C \cite{chaudron1924Wustite570}. 
In its bulk form, the chemical formula can also be expressed as {\wustite}, where ``$x$'' represents the degree of Fe deficiency due to Fe vacancies \cite{bernal2014Wustite,hazen1984Wustite,gavarri2019Wustite}. 
For simplicity, we will subsequently refer to wüstite directly as its stoichiometric form, {\wustiteSlab}, yet recall that it is Fe depleted.
In the context of the interface, when we now introduce a stoichiometric two-layer {\wustiteSlab} slab (complexion) between the Fe and {\magnetite} layers, our DFT calculations align more closely with the experimental results.
The difference between the DFT predictions and the observed plane spacing has decreased from 50\% to less than 5\% with the revised structural model, which includes the {\wustiteSlab} slab.
This improved correlation is convincingly illustrated in the bridging sketch (see \Fig{fig:fig1}{b}-v), which overlaps the experimental charge-density map (see \Fig{fig:fig1}{b}-iv) with the theoretical predictions (see \Fig{fig:fig1}{b}-vi). 
This effectively resolves the discrepancies previously discussed concerning the O and tetrahedral Fe atomic columns highlighted by yellow and white arrows in \Fig{fig:fig1}{b}-iv.

To better understand the energetic reason for the occurrence of this {\wustiteSlab}-type complexion, we further present the relative thermodynamic stability of the different interface structures as a function of the O chemical potential $\Delta \mu_{\text{O}}$ in \Fig{fig:fig1}{c} (see Methods Section~\nameref{Calculations}) \cite{reuter2001DFT,cao2019DFT,meng2021FeFeOint}.
For this purpose, we additionally considered the most stable configurations of the Fe\hkl[001]/{\wustiteSlab}\hkl[001] interface and Fe/{\wustiteSlab}/{\magnetite} heterostructures comprising 1 to 4 intermediate {\wustiteSlab} single atomic layers (see \figs{fig:fig_sites_Fe_FeO,fig:fig_sites_FeO_Fe3O4,fig:fig_charge_FeO_Fe3O4}{}).
Over the whole range of accessible chemical potentials $\Delta \mu_{\text{O}}$, given in terms of the corresponding O$_2$ pressure, the heterostructures are favored according to our analysis. 
In particular, the number of intermediate {\wustiteSlab} layers stabilizing the interface decreases with decreasing pressure or $\Delta \mu_{\text{O}}$. 
More specifically, a transition occurs from 4 down to 2 intermediate {\wustiteSlab} atomic layers at an O chemical potential shift of approximately $-1.7$\,eV.
The pristine Fe/{\magnetite} interface is finally favored below an O chemical potential shift of approximately $-1.9$\,eV, corresponding to an O$_2$ pressure lower than $10^{-19}$\,mbar at 300\,{\degree}C, far below the experimental conditions at which the thin films were grown.

We also emphasize that due to the limited system sizes accessible to the DFT calculations, we did not consider more than 4 intermediate {\wustiteSlab} atomic layers. 
However, given the trend observed in \Fig{fig:fig1}{c}, we expect that the most stable structure in the O-rich region will contain many more {\wustiteSlab} layers. 
A similar effect has been observed recently in DFT calculations of stacked {\wustiteSlab} layers on top of a free Fe\hkl(001) surface \cite{ossowski2023FeOFe}, or for grain boundary complexions in a model bicrystal~\cite{Rickman2013}. 
Also, given the range of accessible chemical potentials in experiments, such extreme cases are not the most relevant ones for the present analysis.

\subsubsection*{The effect of complexion formation on the properties of the Fe/{\magnetite} interfaces} \label{complexion_charge}

\begin{figure}[htbp!]
    \centering
    \includegraphics[trim = 0mm 0mm 0mm 0mm, clip, width=1.0\linewidth]{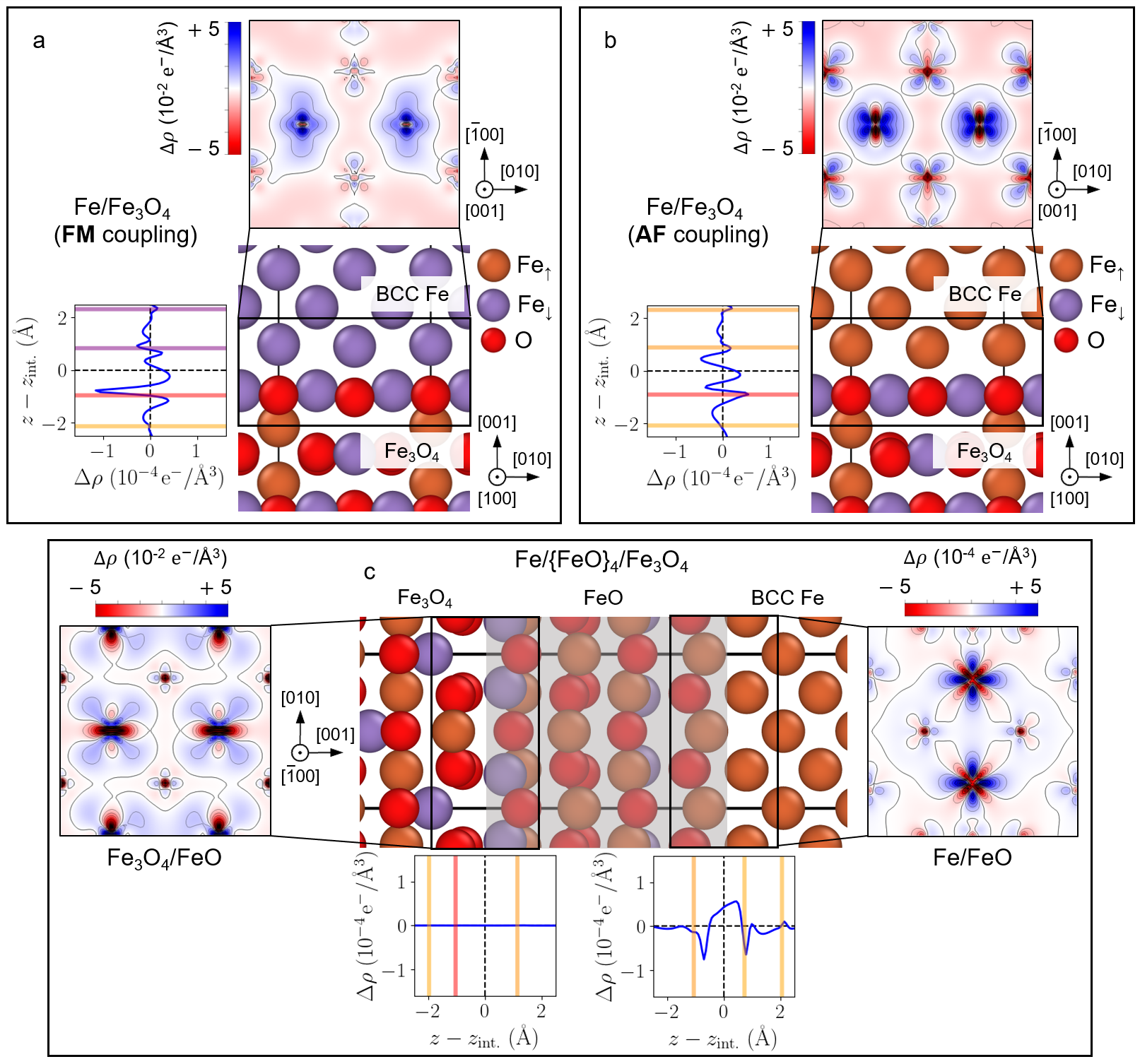}
    \caption{\textbf{Charge transfer at the Fe/{\magnetite} interface and the effect of the presence of FeO-type complexion.} 
    \textbf{a} Charge transfer at the Fe/{\magnetite} interface when the BCC-Fe layer is coupled ferromagnetically to the outermost FeO$_2$-terminated {\magnetite} layer (all Fe spins pointing down), and 
    \textbf{b} with the BCC-Fe layer antiferromagnetically coupled to {\magnetite}.
    \textbf{c} Charge transfer when a 4-layer FeO complexion is located between BCC-Fe and {\magnetite}, both at the FeO/{\magnetite} (left) and the BCC-Fe/FeO (right) interfaces.
    For each subplot is shown the in-plane projection of the charge transfer $\Delta \rho$ in the vicinity of the interface and the line profile in the direction normal to the interface plane as a function of the distance $z-z_{\text{int.}}$ to the interface height $z_{\text{int.}}$. Crystallographic orientations are given with respect to the BCC-Fe lattice.}
    \label{fig:fig2}
\end{figure}

We now investigate in~\Fig{fig:fig2}{} the impact of the formation of such complexions on the charge transfer at the Fe/{\magnetite} interface using DFT calculations. 
As can be inferred from the change in the structure of the whole interface, the transfer properties are also impacted by the presence of {\wustiteSlab} complexions. 
In both cases (with and without a complexion present), most of the charge transfer occurs within the inter-layer region, with electrons coming from the oxide layer.
We also note a stronger electronic transfer at the interface between {\wustiteSlab} and BCC-Fe (\Fig{fig:fig2}{C}, right panel), which is not present at the clean binary Fe/{\magnetite} interface (\Fig{fig:fig2}{A \& B}). 
This is an indication of an increased strength of the whole interface in the case of the Fe/\{FeO\}$_4$/{\magnetite} heterostructure.
This is supported by the stronger predicted work of adhesion of -3.9\,J/m$^2$ between iron and the oxide for the heterostructure compared to -2.7\,J/m$^2$ for the binary interface (see \Tab{tab:values}). 
This is caused by the strong adhesion between BCC-Fe and {\wustiteSlab}, also in line with other DFT studies on the Fe/{\wustiteSlab} interface \cite{ossowski2023FeOFe,meng2021FeFeOint}.

We also investigate the influence of the magnetic coupling between the BCC-Fe and the oxide layer on the electronic transfer, focusing in \Fig{fig:fig2}{A \& B} on the clean binary Fe/{\magnetite} interface.
A ferromagnetic (FM) coupling between the two layers is energetically more favourable. 
However, the energy difference between the FM and antiferromagnetic (AF) couplings amounts to a value of only 0.1\,J/m$^2$, qualifying the two types of magnetic coupling at the interface almost as degenerate states. 
On the other hand, the behavior of the charge transfer at the very interface changes quite noticeably from one magnetic coupling to the other. 
In particular, the negative transfer is more localized with an AF coupling, centered around the position of the outermost octahedral Fe atom of the {\magnetite} layer.

Finally and most importantly, the change in the structure of the Fe/{\magnetite} interface with the oxygen chemical potential, driving the formation of {\wustiteSlab} complexions of various thicknesses, will greatly impact the transport properties of a material containing these defects. 
Indeed, wüstite is inherently off-stoichiometric, with a depleted iron sub-lattice. 
Thus the presence of {\wustiteSlab} layers at the Fe/{\magnetite} interface will act as a sink for iron vacancies. 
These interfaces are omnipresent during redox processes involving iron and its oxides, for instance during hydrogen-based reduction of oxides~\cite{ma2022HBD}, or during the combustion of iron powder~\cite{choisez2022combustion}. 
Thus, an understanding of their structure is of great importance for modeling the transport of atomic species during these processes.

\subsubsection*{Formation of the Fe/{\magnetite} complexions independent of the processing path} \label{process}

\begin{figure}[htbp!]
    \centering
    \includegraphics[trim = 0mm 0mm 3mm 0mm, clip, width=0.85\linewidth]{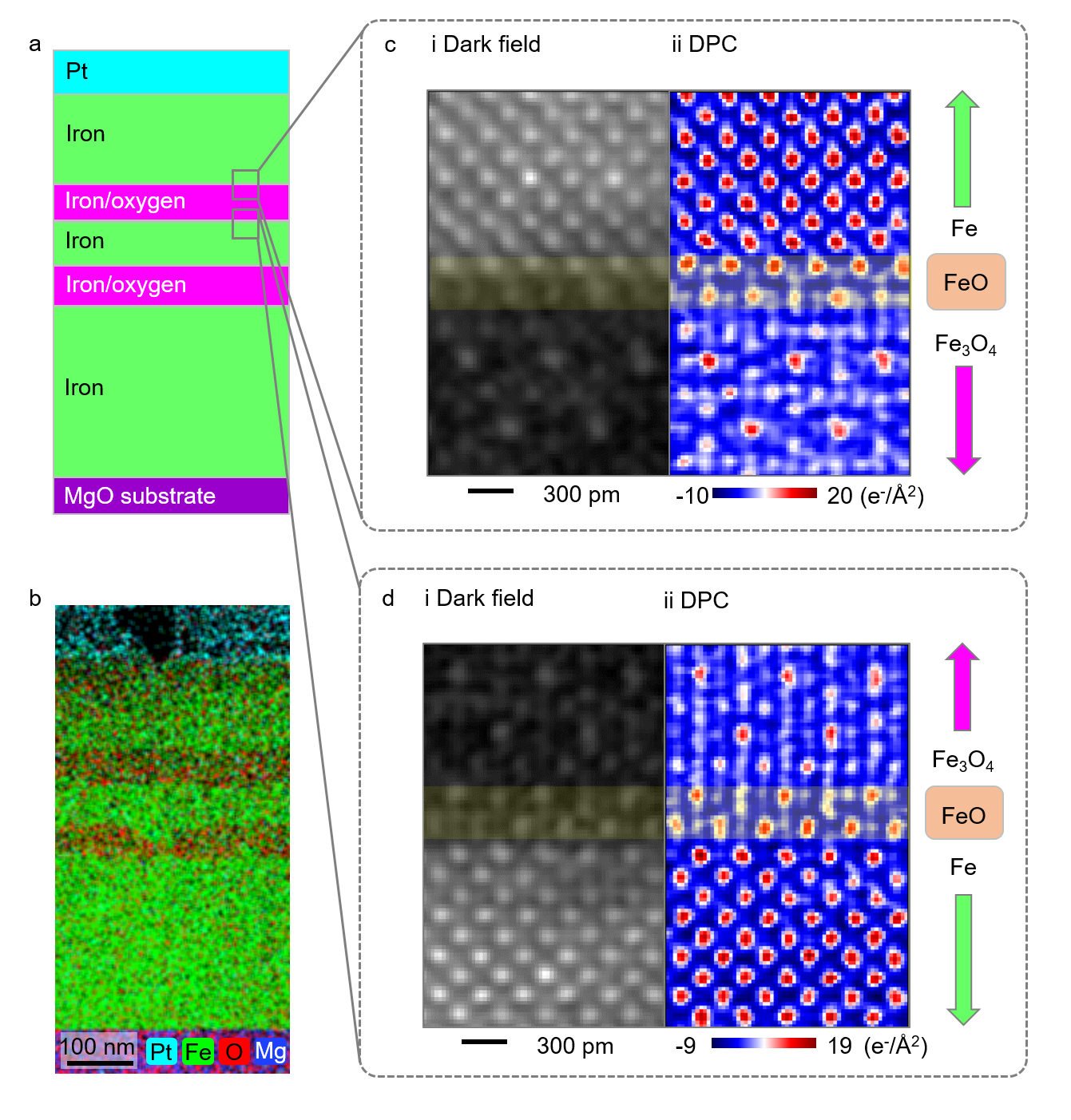}
    \caption{\textbf{The Fe/{\magnetite} complexions form under various fabrication conditions.} 
    \textbf{a} Multi-layer thin film deposition under different controlled external atmospheres, with or without flowing O$_{2}$ gas. 
    \textbf{b} Energy-dispersive X-ray spectroscopy (EDS) map showing the cross-section view of the thin film.
    \textbf{c} Fe is depositing on {\magnetite} under a pure Ar atmosphere, where the {\wustiteSlab}-type complexion is observed at the interface.
    \textbf{d} {\magnetite} is depositing on Fe under a mixed Ar/O$_{2}$ atmosphere, where the {\wustiteSlab}-type complexion is also observed at the interface.
    In \textbf{c} and \textbf{d}, the areas between the red arrows indicate the regions of {\wustiteSlab}-type complexions.
    DPC: differential phase contrast imaging, conducted in a scanning transmission electron microscope.
    }
    \label{fig:fig3}
\end{figure}

Experimentally, we have employed atmospherically controlled PVD to produce the specimens in which we examine the Fe/{\magnetite} interface under varying levels of external O activity (an equivalent measure of chemical potential, controlled by the O$_2$/Ar ratio and pressure). 
This is illustrated in \Fig{fig:fig3}{a} and further demonstrated by energy-dispersive X-ray spectroscopy (EDS) mapping in \Fig{fig:fig3}{b}.
Our aim was to create multi-layer thin films where the Fe/{\magnetite} interfaces occur under different conditions. 
This enables us to assess the general applicability of our observations regarding interfacial structures associated with {\wustiteSlab}. 
In one setup, BCC-Fe was deposited on {\magnetite} under conditions of very low external O activity, as shown in \Fig{fig:fig3}{c} and \fig{fig:DPC_IM}{}. 
In contrast, under a high O activity environment, {\magnetite} was deposited on BCC-Fe, as depicted in \Fig{fig:fig3}{d} and \fig{fig:DPC_MI}{}. 
It should be noted that {\wustiteSlab}-type complexions were observed in both scenarios, as indicated by the arrows in \Figs{fig:fig3}{c \& d}. 

Our multi-layer deposition synthesis approach serves as a method to examine the interface contribution in stabilizing such complexions or any related interfacial phase states.
The thermodynamic discussion and further details on the formation of bulk Fe and {\magnetite} phases during thin film deposition can be found in the Methods Section~\nameref{Calculations} and \figs{fig:Thermo,fig:Supp_TC}{}.
Our observations at the interface between Fe and {\magnetite} involve an interfacial structure closely tied to the {\wustiteSlab} structure. 
In both cases, Fe on {\magnetite} or {\magnetite} on Fe, {\wustiteSlab}-type complexions are stabilized at the interfaces. 
Interestingly, our experimental conditions are based on a synthesis at 300\,{\degree}C under constant external O activity, where {\wustiteSlab}, a high-temperature stable oxide, exists as a interfacial phase under ambient pressure above 570\,{\degree}C \cite{chaudron1924Wustite570}. 
Such observations of {\wustiteSlab}-type complexions occur when the interface evolves towards a complexion-mediated equilibrium state. 
Our findings demonstrate a unique pathway for stabilizing interface phase states that are not stable according to the bulk phase diagram but can be made thermodynamically stable when they are between two adjacent phases, through a deliberately designed interface structure.

\subsubsection*{Diversity of Fe/{\magnetite} interfacial structures} \label{complexion_diversity}

\begin{figure}[!b]
    \centering
    \includegraphics[trim = 1mm 1mm 1mm 1mm, clip, width=1\linewidth]{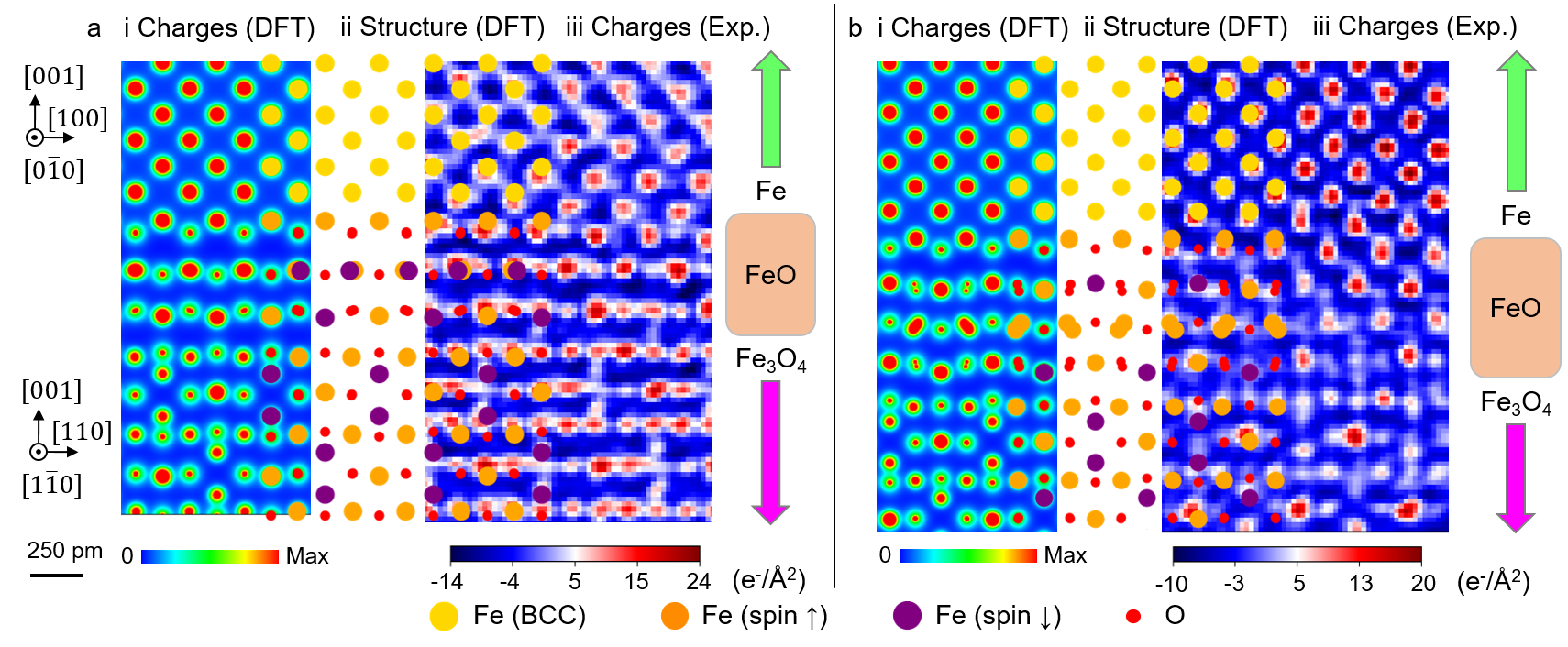}
    \caption{
    \RV{\textbf{Different Regions of the Fe/{\magnetite} interface displaying {\wustiteSlab}-type complexions of various thicknesses.}
    \textbf{i} DFT-computed electronic charge densities, \textbf{ii} DFT-relaxed equilibrium structure, and \textbf{iii} DPC-4DSTEM reconstructed charge density map for \textbf{a} three-layer and \textbf{b} four-layer {\wustiteSlab}-type complexions at the Fe/{\magnetite} interface. 
    The relaxed DFT structure of the interface is presented as a bridging sketch between the two charge density maps in both \textbf{a} and \textbf{b}. 
    DFT: Density functional theory.
    DPC: differential phase contrast imaging, conducted in a 4DSTEM experiment. 
    }
    }
    \label{fig:fig4}
\end{figure}

Throughout our discussion, we have primarily focused on the observation of a two-layer {\wustiteSlab}-type complexion at the Fe\hkl[001]/{\magnetite}\hkl[001] interface.
Our DFT calculations also predict the stability of thicker {\wustiteSlab}-type complexions, as shown in \Fig{fig:fig1}{e} in the O-rich region.
\RV{Additional observations in \Fig{fig:fig4}{}, \figs{fig:DPC_MI_3layer,fig:DPC_IM_4layer}{} corroborate this prediction, showing that in certain regions of the Fe/{\magnetite} interface, the three-layer and four-layer {\wustiteSlab}-type complexion are evident, respectively}. 
Such heterogeneity in complexion thickness may be attributed to a local fluctuation of synthesis conditions leading to a local O$_2$ activity variation.
As seen in \Fig{fig:fig1}{c}, the configurations of the {\wustiteSlab}-type complexion found at the Fe/{\magnetite} interface demonstrate a dependence on local O chemical potential.

It is also worth mentioning that the local strain build-up due to the mismatch between adjacent phases can be another factor in controlling the interface complexions. 
While the lattice mismatch between Fe and {\magnetite} is relatively small, this is not the case with {\wustiteSlab}, with $\varepsilon_{xy}^{\text{BCC-Fe}}=-6.8\,\%$ and $\varepsilon_{xy}^{\text{\wustiteSlab}}=+7.3\,\%$ with respect to the other layer.
This mismatch also favors the formation of defects, such as misfit dislocations and steps, at the Fe/{\magnetite} interface to alleviate strain, examples of which are illustrated in \figs{fig:STEM_IF,fig:DPC_Step,fig:DPC_Semico}{}. 
\RV{For the regions without misfit dislocations, we measured lattice parameters for the Fe phase, the {\wustiteSlab}-type interfacial phase, and the {\magnetite} phase to be approximately 0.29 nm, 0.43 nm, and 0.83 nm, respectively (see black lines in \fig{fig:DPC_Coherent}{}). 
These measurements are consistent with our DFT-predicted values of 0.28 nm, 0.43 nm, and 0.84 nm, respectively, confirming the existence of the interfacial {\wustiteSlab}-type phase.}

In summary, in this study, we employed a combination of advanced characterization techniques and DFT calculations to explore the structural and chemical features at the interfaces between iron and its oxides. 
Experimentally, we observed interfacial {\wustiteSlab}-type complexions at the Fe\hkl[001]/{\magnetite}\hkl[001] interface and used DFT-based thermodynamics to demonstrate their higher stability compared to an interface without presence of such complexions.
Our findings provide an explanation for the formation of complexions at the phase boundaries.
Using DFT calculations, we constructed the complexion diagram of the interface, explaining the origin of the different types of observed interface phase states, depending on the oxygen chemical potential and the adjacent bulk phases.
The atomic structure of the complexions we observed appears in specific phase states, which are significantly different from their equilibrium bulk existence and impact the electrical, mechanical, and transport properties of the interface.
This study thus lays the groundwork for the possible future integration of such structural-chemical complexion states as a material and process design toolbox, accessible to crisp thermodynamics rules.
The specific defect phase state we studied here can leverage profound kinetic, thermodynamic, and mechanical effects that can alter mass and charge transport, phase transformation behavior, interface strength and catalysis, to name but a few features, highlighting the broader principle that thermodynamically stable interface structures can be predicted, designed, implemented and functionally utilized with significant opportunities for advanced materials, processes and interface-related reaction phenomena.


\bmhead{Acknowledgments}

The authors highly appreciate the discussions with Prof. Dr. J{\"o}rg Neugebauer, Prof. Dr. Gerhard Dehm, Prof. Dr. Scheu Christina, Prof. Dr. Gault Baptiste, Dr. Mira Todorova, and Dr. Christoph Freysoldt. 

\section*{Declarations}
\begin{itemize}
\item \textbf{Funding} {
    D.R. acknowledges funding by the European Research Council (ERC) via the ERC advanced grant 101054368. 
    Views and opinions expressed are however those of the authors only and do not necessarily reflect those of the European Union the ERC.
    Neither the European Union nor the granting authority can be held responsible for them.
    X.Z. acknowledge funding by the German Research Foundation (DFG) through the project HE 7225/11-1.
    Work at the Molecular Foundry was supported by the Office of Science, Office of Basic Energy Sciences, of the U.S. Department of Energy under Contract No. DE-AC02-05CH11231.}
\item \textbf{Competing interests} {The authors declare no competing interests}
\item \textbf{Availability of data and materials} {should be addressed to Dr. Xuyang Zhou and Dr. Baptiste Bienvenu}
\item \textbf{Code availability} {
    The Python code used for the electron microscopy analysis in this study is available on GitHub: \href{https://github.com/RhettZhou/pyDPC4D}{https://github.com/RhettZhou/pyDPC4D}.
    The Python code used for the Thermo-Calc study is also available on GitHub:\\ \href{https://github.com/YXWU2014/IronOxide_TC}{https://github.com/YXWU2014/IronOxide\_TC}
    }
\item \textbf{Authors' contributions} {
    D.R. and X.Z. secured funding. 
    D.R. conceived of the presented idea and supervised the project. 
    X.Z. conducted the experimental study.
    B.B. performed the atomistic simulations. 
    Y.W. and A.K. executed the thermodynamic calculations. 
    C.O. contributed to the electron microscopy data analysis. 
    X.Z., B.B., and Y.W. wrote the original paper. 
    All the authors revised the paper.
    }
\end{itemize}

\begin{appendices}

\section*{Materials and Methods}\label{method}

\subsection*{Synthesis}\label{thin_film_synthesis}
Five-layered Fe/iron oxide thin films were prepared in a BesTeck PVD cluster (MPIE, Düsseldorf, Germany) by direct current (DC) sputtering a pure Fe target (99.99\,\%, Mateck, Germany) with a power of 150\,W in the alternated Ar and O$_{2}$ gas environment at 300\,{\degree}C.
The chamber was pumped to a base pressure of 5.0 × 10$^{-8}$\,mbar before the sputtering experiments. 
Subsequently, the Fe thin film layer was first deposited on a \hkl[001] textured single-crystalline MgO substrate at a pressure of 5 × 10$^{-3}$\,mbar and an Ar flux of 40\,standard cubic centimeters per minute (SCCM). 
Then an extra O$_{2}$ flux of 2\,SCCM was introduced for sputtering the iron oxide thin film layer.
The shutters for both the Fe target and the substrate were closed during the transition of the sputtering gas environment.
The deposition rate was kept at approximately 1.9\,nm/min. 
By modulating the gas environment, we obtained a multi-layer thin film with a total thickness of approximately 260\,nm, see \fig{fig:EDS}{a}.
After sputter deposition, the thin film was then cooled to room temperature in the vacuum chamber.

\subsection*{Characterization}\label{Characterization}
\subsubsection*{Electron microscopy}\label{TEM}
\begin{figure}[!htb]
    \centering
    \includegraphics[trim = 0mm 0mm 0mm 0mm, clip, width=1\linewidth]{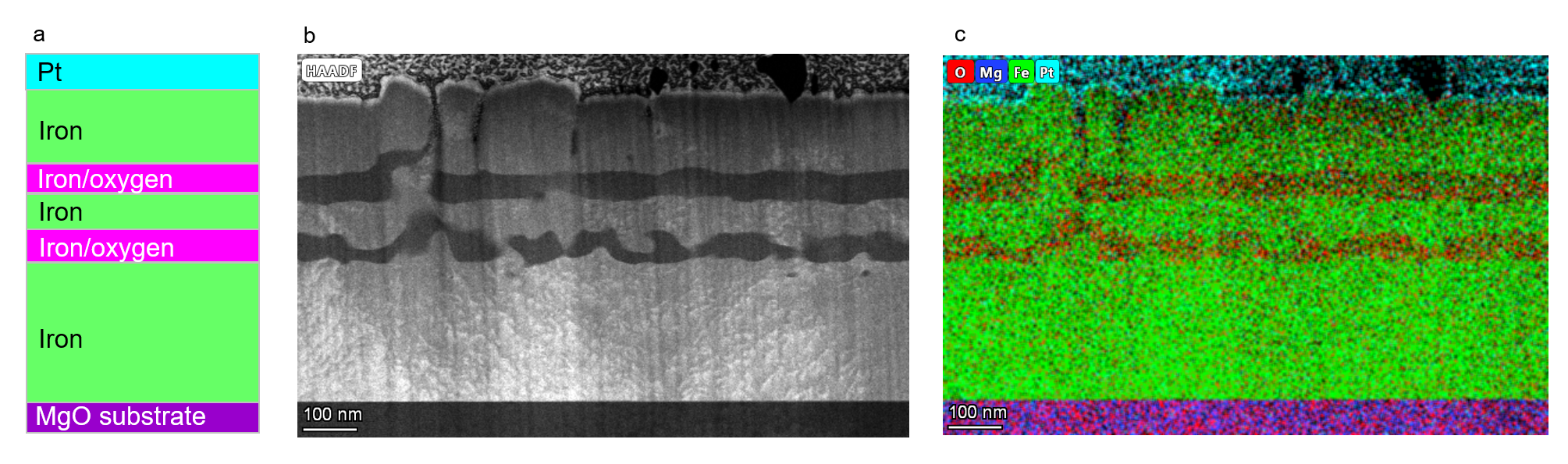}
    \caption{\textbf{Overview of the multi-layer thin film.} 
    \textbf{a} An illustrative representation of the structure; 
    \textbf{b} STEM - High-angle annular dark field (HAADF) image and \textbf{c} EDS map showing the cross-section view of the thin film.}
    \label{fig:EDS}
\end{figure}

We prepared the transmission electron microscopy (TEM) lamellae using the lift-out procedure in an FEI Helios Nanolab 600i focused ion beam (FIB) dual-beam microscope.
Initially, the TEM lamellae were thinned to a thickness of less than 50\,nm using an accelerating voltage of 30\,kV. 
Subsequently, a careful polishing step was performed at an accelerating voltage of 5\,kV.
The thinnest region of the TEM lamellae can reach a thickness of less than 10\,nm.
The STEM-HAADF images and EDS data presented in \figs{fig:EDS}{b\&c} were acquired using a Cs probe-corrected FEI Titan Themis 60-300 microscope operating at 300\,kV. 
The semi-convergence angle was set to 23.6\,mrad, and the semi-collection angle ranged from 78 to 200\,mrad.
The EDS data was collected by acquiring a series of at least 500 frames, with each probe dwell time recorded for a duration of 20\,$\mu$s.
The interface between Fe and iron oxide is sharp, but exhibits a faceted morphology that can be seen in the STEM-HAADF image (see \fig{fig:EDS}{b}) and the EDS mapping (see \fig{fig:EDS}{c}).
In this work, we specifically study the defect-free and strain-free interface to isolate factors influencing complexion formation. 
This clean interface structure (see \Fig{fig:fig1}{} and \Fig{fig:fig3}{}) not only facilitates accurate high-resolution imaging and DPC measurements but also enables direct comparisons with DFT calculations to provide deep insights into the atomic mechanisms at play.

\begin{figure}[!htb]
    \centering
    \includegraphics[trim = 0mm 0mm 0mm 0mm, clip, width=1\linewidth]{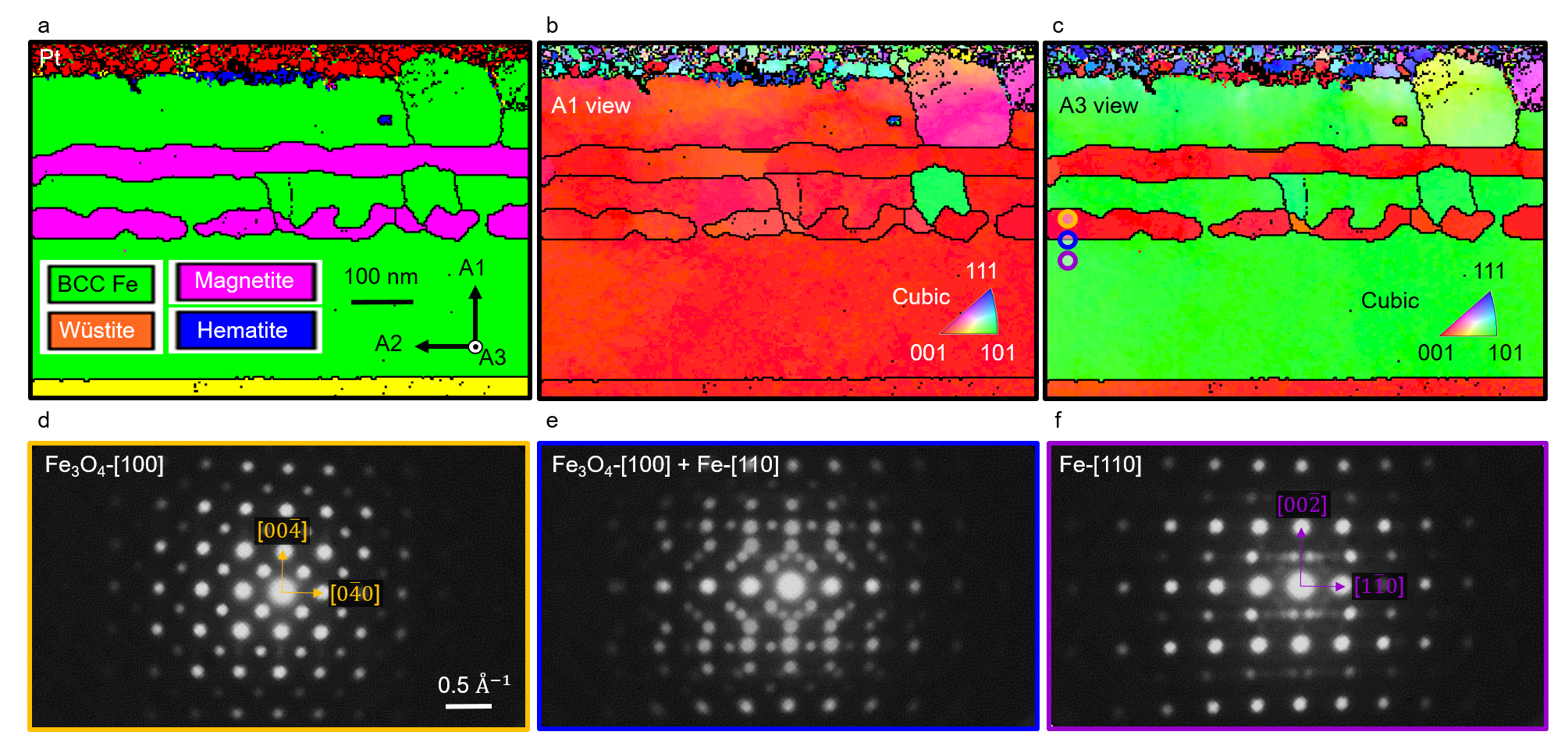}
    \caption{\textbf{Structural characterization of the multi-layer thin film.} 
    \textbf{a} Phase map and orientation maps obtained from \textbf{b} View A1 and \textbf{c} View A3, reconstructed from the precession-assisted 4DSTEM datasets of the cross-sectional view of the thin film. 
    The sample coordinates A1-A3 were defined in the phase map.
    \textbf{d-f} Diffraction patterns from regions highlighted in \textbf{c} (orange, blue, and purple circles) are \textbf{d} magnetite-\hkl[100], \textbf{e} the interface between magnetite-\hkl[100] and Fe-\hkl[110], and \textbf{f} Fe-\hkl[110]. 
    Additional diffraction spots in \textbf{d} and \textbf{f}, not part of the Fe or {\magnetite} lattice, arise from the native surface oxide layer on Fe.
    }
    \label{fig:4DSTEM}
\end{figure}

We analyzed the phase and orientation of the multi-layer thin films using the precession-assisted 4DSTEM technique \cite{jeong20214D}, as shown in \fig{fig:4DSTEM}{}.
The 4DSTEM data sets were acquired using the TemCam-XF416 pixelated complementary metal-oxide-semiconductor detector (TVIPS) installed in a JEM-2200FS TEM (JEOL) operating at 200\,kV.
To create a quasi-kinematic diffraction pattern, we precessed the incident electron beam by 0.5\,{\degree} while scanning the specimens with a step size of 3\,nm.
The collected 4DSTEM data set was then indexed using the ASTAR INDEX program, and the phase and orientation were mapped using the TSL OIM Analysis 8 software package.
Structural characterization in \fig{fig:4DSTEM}{a} reveals that the dominant phases for the Fe layers and the oxide layers are the BCC phase and the inverse spinel phase, respectively.
The primary out-of-plane growth directions for both phases appear to be \hkl[001] (see \fig{fig:4DSTEM}{b}), suggesting an epitaxial growth mechanism. 
\fig{fig:4DSTEM}{c} presents the in-plane orientation between different layers.
A 45\,{\degree} rotation has been observed between the BCC-Fe and the {\magnetite} layers, indicating an orientation relationship matching the one determined previously by Davenport {\etal} \cite{davenport2000MagFeOri}, namely $\text{Fe}\hkl[001]_z \parallel \text{\magnetite}\hkl[001]_z$, $\text{Fe}\hkl[100]_x \parallel \text{\magnetite}\hkl[110]_x$.

The DPC-4DSTEM data sets were acquired using the electron microscope pixel array detector (EMPAD) equipped in the aforementioned Titan microscope at 300\,kV.
For each probe position, the EMPAD captured the convergent beam electron diffraction (CBED) pattern at a semi-convergence angle of 23.6\,mrad, with an exposure time of 1\,ms per frame.
All CBED patterns had a uniform size of 128 × 128\,$\text{pixel}^2$ with a pixel size of 2.0\,mrad. 
The scanning step size was selected within the range of 13 to 25\,pm.
A smaller scanning step size offers a higher spatial resolution for accurately resolving the positions of atomic columns. 
This is demonstrated in \fig{fig:DPC_M13}{}, which showcases the experimental DPC-4DSTEM reconstruction of {\magnetite} oriented in the \hkl[110] direction. 
However, it is important to note that a smaller scanning step size can introduce distortions due to sample drifting.
On the other hand, scanning with a larger step size can reduce image distortion, but it may compromise the resolution required to resolve lighter O atomic columns. 
After optimization, we determined that the optimal scanning step size for the current experiments is 18\,pm, as depicted in \fig{fig:DPC_M18}{}.
Annular bright field (ABF)-STEM imaging concurrently provides contrast for both heavy Fe atomic columns and light O atomic columns \RV{\cite{saxton1978ABF,okunishi2009ABF,Chen2018Twin,Jiang2021If,Gao2024Fe3O4}}. 
However, ABF-STEM phase contrast measurements are challenging due to complex contrast transfer functions \cite{ophus2023ABF}.
In this study, we consistently employed DPC-4STEM as the primary method for imaging the interface structure \RV{as we find the multi-faceted information it provides to be valuable to help us address our scientific questions effectively}.

We reconstructed the DPC-4DSTEM data sets using the in-house developed Python script pyDPC4D (GitHub link: \href{https://github.com/RhettZhou/pyDPC4D}{https://github.com/RhettZhou/pyDPC4D}).
The reconstructed data provide information including the virtual dark-field image, the center of mass of the transmitted beam, the electric field vector map, the projected electrostatic potential map, and the charge-density map, see \fig{fig:DPC_M18}{}.
Here we will briefly review the key steps of the data reconstruction.
More details regarding the data reconstruction are referred to our previous publication \cite{zhou2023FeBC}.

The DPC-4DSTEM data set consists of a two-dimensional (2D) grid of probe positions in real space along with corresponding 2D diffraction patterns in reciprocal space. \cite{ophus20194D,savitzky2021Py4DSTEM}. 
Each recorded diffraction pattern exhibits a bright field disk that contains information about the momentum transfer of the electron beam \cite{shibata2012DPC,muller2014DPC,muller2017DPC,hachtel2018DPC,gao2019DPC}.
By analyzing the displacement of the center of mass of the bright field disk, we can track the momentum transfer of the electron beam as a function of the probe position. 
According to the Ehrenfest theorem \cite{ehrenfest1927DPC}, the electric field of a thin specimen is related to the momentum transfer of the electrons in the beam \cite{pennycook2011STEM}.
The integration of the electric field yields the projected electrostatic potential \cite{muller2014DPC,muller2017DPC,hachtel2018DPC}.
The charge-density (including electrons and protons) is proportional to the divergence of the electric field as dictated by Gauss's law \cite{muller2014DPC,muller2017DPC,gao2019DPC}.

We conducted in-situ heating experiments to examine the stability of the \wustiteSlab-type complexion. 
Using the lift-out procedure with an FEI Helios Nanolab 600i FIB dual-beam microscope, we prepared the TEM lamellae on a DENSsolutions chip. 
We then performed electron microscopy experiments in the Cs probe-corrected FEI Titan Themis 60-300 microscope using similar operating parameters as previously mentioned. 
The experimental conditions included \SI{2}{\hour} of in-situ TEM heating at \SI{300}{{\degree}C}, with a heating and cooling rate of \SI{5}{{\degree}C/\second}. 
Overall, the structure remained unchanged after prolonged heating; although we observed some local reorganization through a ledge growth mechanism at the atomic scale, the \wustiteSlab-type complexion consistently appeared at the interface (see \figs{fig:in-situ,fig:DPC_MI_3layer}{}).

For quantifying local composition and valence state transitions across the interface, we used a Gatan Quantum spectrometer to acquire electron energy loss spectroscopy (EELS) data at \SI{300}{kV} with an entrance aperture of \SI{35}{mrad}.
We estimated the atomic ratio of Fe to O by integrating the intensities of the Fe-L3 and O-K edges.
Additionally, we performed peak decomposition on the Fe-$L_{3}$ edges to determine the fractions of various oxidation states \cite{van2002EELS}.
As \fig{fig:EELS}{} illustrates, EELS analysis provides insights into the O atomic ratio and valence state transitions across the interface of the specimen heated in-situ at \SI{300}{{\degree}C} for \SI{2}{\hour}. 
Although the complexion region appears in HAADF images, a gradient in valence states at the interface, rather than a sharp transition, occurs due to beam broadening and overlapping signals of Fe, $Fe^{2+}$, and $Fe^{3+}$ in {\magnetite}, complicating the direct observation of complexions using EELS.

\subsubsection*{Multi-slice image simulation}
We generated a synthetic 4D-STEM data set using the $\mu$STEM simulation suite, which employs the multislice method \cite{allen2015Mustem}.
By comparing the experimental and simulated CBED patterns (see \fig{fig:DP_Thickness}{}), we estimated the specimen thickness to be approximately 8.3\,nm. 
We analyzed the simulated data set following the same procedure as that for the experimental data in the DPC-4DSTEM analysis \cite{zhou2023FeBC}.
The simulation setup was intended to mimic the parameters used during experimental data acquisition, such as an accelerating voltage of 300\,kV, a semi-convergence angle of 23.6 mrad, a sample thickness of 8.3\,nm, and a probe spacing of 13\,pm.
\fig{fig:mSTEM_M}{} displays the reconstruction for {\magnetite} oriented in the \hkl[110] direction. 
To elucidate the origin of the contrast observed in the charge-density map, we conducted three series of simulations, varying parameters such as sample thickness, focus, and probe spacing, as illustrated in \fig{fig:mSTEM_Para}{}.

\subsection*{Computational Details}\label{Calculations}

\subsubsection*{DFT calculations}
All \textit{ab initio} calculations reported in this work were performed within DFT, as implemented in the \textsc{Vasp} code \cite{Kresse1996}. 
Projector augmented wave (PAW) pseudopotentials including 8 and 6 valence electrons were used to model Fe and O atoms respectively, with the exchange-correlation potential approximated using the GGA-PBE functional \cite{Perdew1996}.
A plane-wave basis at a cutoff energy of 500\,eV was used, and the Brillouin zone was sampled by a $\Gamma$-centered $k$-point mesh of density 0.015\,\AA$^{-1}$.

Magnetism is treated within the collinear approximation. 
\RV{For the BCC-Fe and {\magnetite} layers, all magnetic moments are initialized as in the ground-state magnetic order of each structure, namely ferromagnetic (FM) for BCC-Fe and ferrimagnetic (FeM) for {\magnetite}}. 
\RV{As for the {\wustiteSlab}-type complexions, we search - by using DFT - for the lowest energy magnetic ordering state by considering different arrangement of the Fe magnetic moments within the complexion layer. 
We did so since it might well be that the bulk AF order of FeO is not the most energetically favorable, given the thinness of the complexions. Only the lowest energy configurations are presented, and all values derived from these structures, which all differ from the bulk AF magnetic order of {\wustiteSlab}.}

Structural parameters (lattice parameter and bulk modulus) for the three materials considered here (BCC-Fe, {\wustiteSlab} and {\magnetite}) are presented in \Tab{tab:tab_DFT_bulk}. 
Are presented both standard DFT (GGA-PBE functional) and DFT$+U$ values, compared with experimental references. 
It is worth noting that standard DFT fails to correctly account for the electronic correlations found among iron oxides, therefore usually motivating the use of a Hubbard $U$ term on $3d$-orbitals of Fe atoms. 
However, such a correction yields a poor description of pure Fe (as can be seen in \Tab{tab:tab_DFT_bulk}) for structural parameters, but also the relative stability between different crystal structures and magnetic orders. 
Since both pure Fe and iron oxides coexist in all interface structures studied here, we chose to use a consistent description of Fe atoms across all three materials, therefore opting for standard DFT, without Hubbard $U$ correction.
This choice yields structural properties in good agreement with experimental data for all three materials (see \Tab{tab:tab_DFT_bulk}).

Charge-density plots (see for instance \Figs{fig:fig1}{b\&d}) were obtained from ``all-electron'' DFT calculations in the sense of the PAW method, {\ie} as the sum of the self-consistent valence electron density and the core density contained in the PAW pseudopotentials.

Electron transfer at the interface, presented in \Fig{fig:fig2}{}, is obtained through the DFT-computed electronic charge density difference $\Delta \rho$ defined as
\begin{equation}
    \Delta \rho = \rho_{\text{int.}}-\sum_i{\rho_{\text{slab, }i}}
\end{equation}
where $\rho_{\text{int.}}$ is the charge density of the whole interface, and $\rho_{\text{slab, }i}$ are the charge densities of the different slabs contained in the interface model, {\ie} for the binary BCC-Fe/{\magnetite} the two BCC-Fe and the {\magnetite} slabs, and for the heterostructure, the BCC-Fe, the FeO layers as a whole and the {\magnetite} slabs. The projection in the $xy$-plane is obtained by integrating the charge density difference in the volume contained across the area of the interface, with $\pm 2\,${\AA} height from the location of the interface. Line profiles were obtained by integrating the charge density difference in the $xy$-plane at different heights along the direction normal to the interface plane.

\subsubsection*{Interface models}

All interface models are constructed with periodic boundary conditions along all three Cartesian directions. 
They contain stackings of 7 layers of BCC-Fe\hkl(100) planes, 9 layers of {\magnetite}\hkl(100) planes, and up to 5 layers of \hkl(100) {\wustiteSlab} planes. 
Initial lattice mismatches between the different materials in all interfaces considered are presented in \Tab{tab:values}. 
When constructing the interfaces, the in-plane lattice parameter is fixed, while the lattice parameter is kept to its bulk value for each slab along the direction normal to the interface plane.

For all interface configurations, we then start by varying the relative position of the two slabs specifically among the high-symmetry sites.
The adhesion energy $\gamma_{\text{adh}}$ between the two slabs of the interface is then determined as:
\begin{equation}
    \gamma_{\text{adh}}=\dfrac{E^{\text{tot}}-\sum_{i} E^{\text{slab}}_i}{2\,S},
    \end{equation}
where $E^{\text{tot}}$ is the total energy of the interface model, $E^{\text{slab}}_i$ are energies of the individual slabs in vacuum in vacuum, and $S$ is the in-plane area of the interface. 
Using this definition, a negative adhesion indicates a stable interface between slabs in presence, and allows to find its most stable configuration.

For each site, $\gamma_{\text{adh}}$ is then evaluated as a function of the interface separation distance $d_{\text{surf}}$ without relaxing atomic positions (see for instance \fig{fig:fig_sites_Fe_Fe3O4}{} for the case of the Fe\hkl[001]/{\magnetite}\hkl[001] interface). 
For the most stable site among the ones considered, the atomic positions are then allowed to relax until the remaining forces on all atoms in all three Cartesian directions are less than 5\,meV/{\AA}. 
The geometry of the cell is finally also allowed to relax until the remaining stresses are close to zero. 
The strains after relaxation are given for each layer in \Tab{tab:values} for all interfaces considered.

The structure presented in \Fig{fig:fig1}{d}-i, and all Fe/{\wustiteSlab}/{\magnetite} heterostructures, are obtained by matching the most stable relative positions of the three BCC-Fe, {\wustiteSlab} and {\magnetite} layers for each single interface structure, {\ie} between Fe/{\wustiteSlab} and {\wustiteSlab}/{\magnetite}. 
Looking at the heterostructures, this results in a slight change of the most stable position of the BCC-Fe slab with respect to the {\magnetite} slab as compared to the Fe/{\magnetite} interface of \Fig{fig:fig1}{b}-ii. This can be rationalized considering the proximity between the two {\wustiteSlab} and {\magnetite} structures. 
Indeed, they share the same FCC-O sublattice, while the Fe atoms fill different interstitial sites for the two oxide structures. 
Here, the structure of the interface presented in \Fig{fig:fig1}{d}-i, which is the most stable configuration, corresponds to the position of the {\wustiteSlab} layer with respect to the {\magnetite} layer where these two FCC-O sublattices match.

\subsubsection*{Complexion diagram}

The complexion diagram presented in \Fig{fig:fig1}{e} is constructed as detailed in Refs. \cite{reuter2001DFT,cao2019DFT}. 
For this purpose, we need the interface energy for each structure of interest, computed using DFT. Those are Fe/{\magnetite}, Fe/{\wustiteSlab} and Fe/{(\wustiteSlab)}$_{n}$/{\magnetite}, with $n$ the number of intermediate {\wustiteSlab} layers intercalated between the BCC-Fe and {\magnetite} slabs.

To study the stability of each of these structures, we model the experimental conditions as the following: a clean Fe\hkl(001) surface (deposited on the MgO substrate) is then covered with Fe and O atoms coming from reservoirs of given chemical potentials $\mu_i$.
The chemical potentials of each species are given by $\mu_{\text{Fe}}=E_{\text{Fe}}^{\text{BCC}}+\Delta \mu_{\text{Fe}}$ and $\mu_{\text{O}}=\sfrac{1}{2}E_{\text{O$_2$}}^{\text{O$_2$}}+\Delta \mu_{\text{O}}$, where $E_{\text{Fe}}^{\text{BCC}}$ and $E_{\text{O$_2$}}^{\text{O$_2$}}$ are the DFT energies of BCC-Fe and the O$_2$ molecule respectively, while $\Delta \mu_{\text{Fe}}$ and $\Delta \mu_{\text{O}}$ are the changes in their chemical potentials with respect to these two reference states. 
The relative energies $\gamma$ of the different interface configurations, defined per surface area, are then given by:
\begin{equation}
    2\,S\,\gamma=\left[ E^{\text{tot}}_{\text{int.}} - E_{\text{ref.}} - n_{\text{Fe}}\,E_{\text{Fe}}^{\text{BCC}} - \dfrac{1}{2}n_{\text{O}}\,E_{\text{O$_2$}}^{\text{O$_2$}} \right] - n_{\text{Fe}}\,\Delta \mu_{\text{Fe}} - n_{\text{O}}\,\Delta \mu_{\text{O}}
\end{equation}\label{eq:interface_phases}
with $E^{\text{tot}}_{\text{int.}}$ the total energy of the interface, $E_{\text{ref.}}$ the energy of the reference (here a clean Fe\hkl(001) surface), $n_{\text{Fe}}$ and $n_{\text{O}}$ the number of Fe and O atoms contained in the deposited slabs ({\ie} excluding the reference Fe\hkl(001) surface). For the analysis presented in \Fig{fig:fig1}{e}, we set $\Delta \mu_{\text{Fe}}=0$ and let the change in O chemical potential $\Delta \mu_{\text{O}}$ be the varying parameter.
Also, the in-plane lattice parameter of each material (BCC-Fe, {\wustiteSlab} and {\magnetite}) is set to the DFT-computed one of BCC-Fe, to mimic the experimental conditions where the oxide layer is deposited on top of a BCC-Fe layer. 
The same complexion diagram was also computed using the lattice constant of {\magnetite} to mimic the other experimental condition. 
A third set of calculations was also carried using the average lattice parameter of the three materials, weighed by their respective bulk moduli. 
The exact same trend ({\ie} transition from a 4-layer FeO complexion to the clean Fe/{\magnetite} interface with decreasing oxygen chemical potential) was observed under all three conditions.

\subsubsection*{Bulk thermodynamics}\label{Thermo_calculations}

The Fe/{\magnetite} multilayer thin film is created by atmospherically controlled PVD under varying levels of external O activity, and this is compared with bulk thermodynamics, which serves as a guide for thin film deposition. It should be emphasized that the bulk thermodynamics conducted here are not aimed at explaining the kinetic path of interface complexion formation, but rather at aiding the understanding of bulk phase formation during thin film deposition.

We compared two types of Gibbs energy evaluations. The first type involves full equilibrium calculations, mapping out stable phases as functions of both temperature and O activity. 
\fig{fig:Thermo}{a} shows the thermodynamic equilibrium as a function of O activity and temperature. 
At high O activity, this equilibrium favors {\hematite} (hematite). Notably, {\wustiteSlab} is absent across the neighboring regimes of the synthesis conditions illustrated in \fig{fig:Thermo}{a}.

The second type focuses on the instantaneous driving force at the onset of the reaction, which represents the energetic shift in the system when an infinitesimal amount of solid phase (either Fe or iron oxide) is deposited from a supersaturated gas phase \cite{hillert1999Driving}. 
A metastable phase is often frozen during low-temperature deposition due to the limited diffusivity for further phase transformation.
Therefore, we computed the instantaneous chemical driving force for reaction onset for iron oxide polymorphs and BCC-Fe, in reference to depositing from the gas phase, as a function of temperature and O activity.

In \fig{fig:Thermo}{b}, we labeled the plots with O activities corresponding to the experimental conditions: 1) BCC-Fe-forming condition at an infinitesimal O activity at 300\,{\degree}C; 2) an experimental magnetite-forming condition, which corresponds to a volumetric flow rate ratio of \(40/2\) for Ar and O at 300\,{\degree}C; 3) an experimental {\hematite}-forming condition, which corresponds to a volumetric flow rate ratio of \(40/10\) for argon and O at 25\,{\degree}C. 
All PVD experiments were conducted at a pressure of \(0.5 \, \text{Pa}\) (\(5 \times 10^{-3} \, \text{mbar}\)). 
We observed a consistent match between the phases synthesized under specific conditions (see \Figs{fig:fig3}{c \& d}) and the thermodynamic predictions derived from calculating the maximum instantaneous chemical driving force.

We have used the Gibbs energy assessments of BCC-Fe from the TCFE13 database, as well as {\hematite}, {\magnetite}, and {\wustiteSlab}, and the gas phase, from the SSUB5 database of Thermo-Calc. 
Additional details can be found in the GitHub repository for these calculations:\\ \href{https://github.com/YXWU2014/IronOxide_TC}{https://github.com/YXWU2014/IronOxide\_TC}.

\clearpage

\subsection*{Supplemental Tables}\label{ST}

\begin{center}
    \begin{table}[!htb]
        \caption{Lattice constants ($a_0=a=b=c$, in {\AA}) and bulk modulus $B_0$ (in GPa) of cubic BCC-Fe, {\wustiteSlab} and {\magnetite} obtained using DFT (GGA-PBE) and $\text{DFT}+U$ (GGA-PBE with $U_{\text{Fe}}=4\,\text{eV}$) compared to experimental data from indicated references.}
        \label{tab:tab_DFT_bulk}
        \centering
        \setlength{\tabcolsep}{4pt} 
        \begin{tabular}{l c c|c c|c c}
             & \multicolumn{2}{c}{DFT} & \multicolumn{2}{c}{$\text{DFT}+U_{\text{Fe}}$} & \multicolumn{2}{c}{Expt.} \\
            Material & $a_0$ & $B_0$ & $a_0$ & $B_0$ & $a_0$ & $B_0$ \\
            \hline
            Fe (BCC/FM) & $2.83$ & 188 & $2.95$ & 123 & $2.87$ \cite{Kittel1966} & 164 \cite{Dorogokupets2017_iron} \\
            \hline
            FeO (NaCl/AF) & $4.30$ & 172 & $4.34$ & 164 & $4.30 - 4.33$ \cite{McCammon1984_wustite} & $150 - 180$ \cite{McCammon1984_wustite} \\
            \hline
            {\magnetite} (spinel/FeM) & $8.40$ & 172 & $8.47$ & 191 & $8.40$ \cite{Haavik2000_Fe3O4} & 183 \cite{Haavik2000_Fe3O4} \\
            \hline
        \end{tabular}
    \end{table}
\end{center}

\begin{center}
    \begin{table}[!htb]
        \caption{In-plane lattice mismatch $\varepsilon_x$ and $\varepsilon_y$ (in $\%$) along the $X$ and $Y$ axis respectively, and adhesion energy $\gamma_{\text{adh}}$ (in J/m$^2$). All properties are reported for both the initial (upper values) and the fully relaxed geometries, {\ie} atomic positions and the geometry of the cell (lower values).}
        \label{tab:values}
        \centering
        \setlength{\tabcolsep}{4pt} 
        \begin{tabular}{l|c c c|c}
             & \multicolumn{3}{c}{$\left( \varepsilon_x,\,\varepsilon_y \right)$} &  \\
            Interface & Fe & FeO & Fe$_3$O$_4$ & $\gamma_{\text{adh}}$ \\
            \hline
            \multirow{2}{*}{Fe/Fe$_3$O$_4$} & $\left( +2.4,\,+2.4 \right)$ & \multirow{2}{*}{/} & $\left( -2.3,\,-2.3 \right)$ & $-2.58$ \\
             & $\left( +2.9,\,+2.8 \right)$ &  & $\left( -1.8,\,-1.9 \right)$ & $-2.96$ \\
            \hline
            \multirow{2}{*}{Fe/FeO} & $\left( +3.6,\,+3.6 \right)$ & $\left( -3.4,\,-3.4 \right)$ & \multirow{2}{*}{/} & $-2.11$ \\
             & $\left( +3.2,\,-0.5 \right)$ & $\left( -3.8,\,-7.3 \right)$ &  & $-3.68$ \\
            \hline
            \multirow{2}{*}{FeO/Fe$_3$O$_4$} & \multirow{2}{*}{/} & $\left( -1.2,\,-1.2 \right)$ & $\left( +1.2,\,+1.2 \right)$ & $-1.76$ \\
             &  & $\left( +0.3,\,-3.5 \right)$ & $\left( +2.8,\,-1.1 \right)$ & $-2.49$ \\
            \hline
            \multirow{2}{*}{Fe/$\{\text{FeO}\}_2$/Fe$_3$O$_4$} & $\left( +4.0,\,+4.0 \right)$ & $\left( -3.1,\,-3.1 \right)$ & $\left( -0.7,\,-0.7 \right)$ & $-3.74$ \\
             & $\left( +4.7,\,+2.7 \right)$ & $\left( -2.4,\,-4.3 \right)$ & $\left( -0.1,\,-1.9 \right)$ & / \\
            \hline
        \end{tabular}
    \end{table}
\end{center}

\clearpage

\subsection*{Supplemental Figures}\label{SI}

\begin{figure}[!htb]
    \centering
    \includegraphics[trim = 0mm 0mm 0mm 0mm, clip, width=1\linewidth]{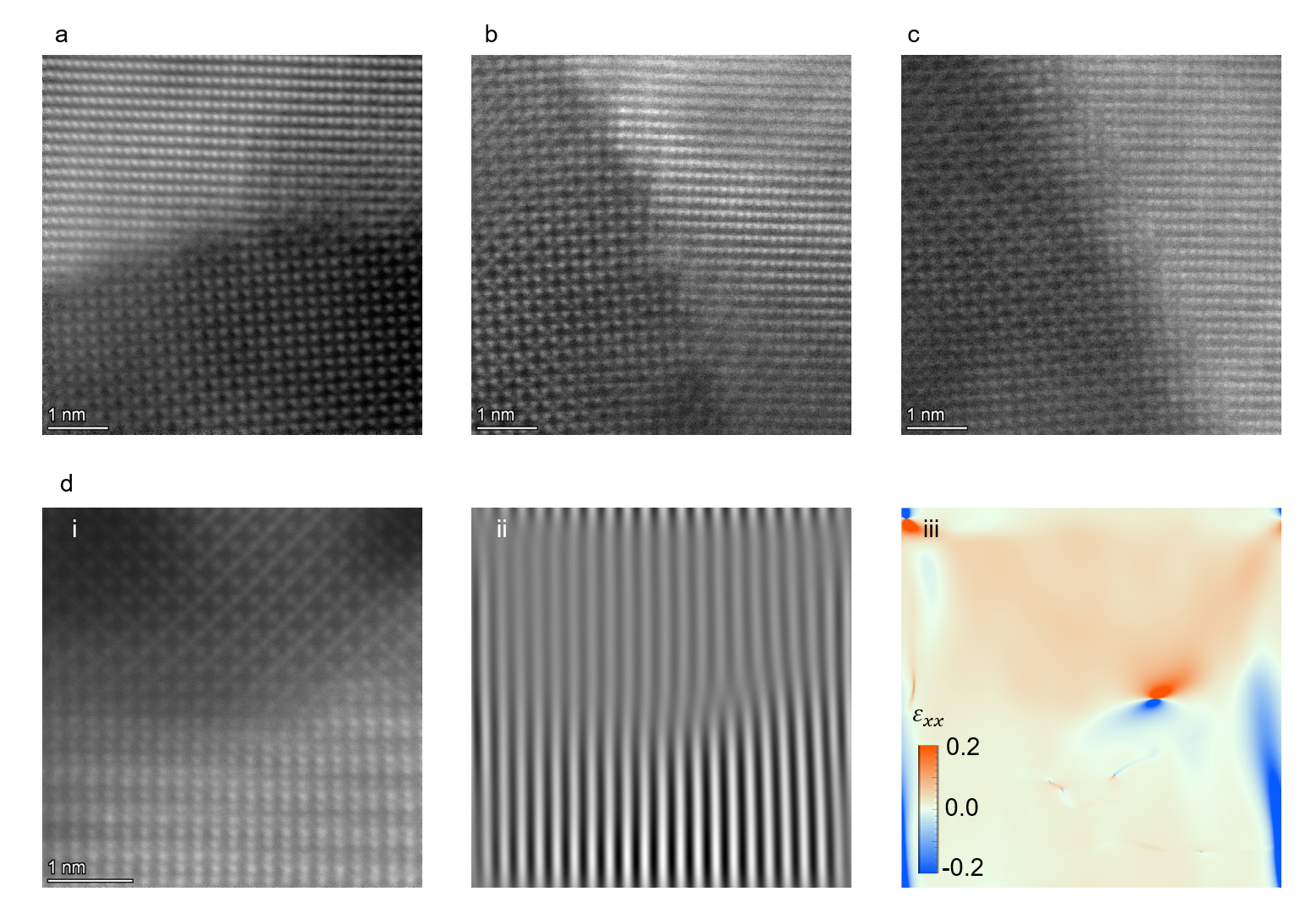}
    \caption{\textbf{Atomic structural characterization of the multi-layer thin film.} 
    \textbf{a-d} display high-resolution STEM-HAADF images of the interface between Fe and {\magnetite}. 
    \textbf{d-ii-iii} further present the Bragg-filtered phase map and strain map for the same regions as shown in \textbf{d-i}.}
    \label{fig:STEM_IF}
\end{figure}

\begin{figure}[!htb]
    \centering
    \includegraphics[trim = 0mm 0mm 0mm 0mm, clip, width=1\linewidth]{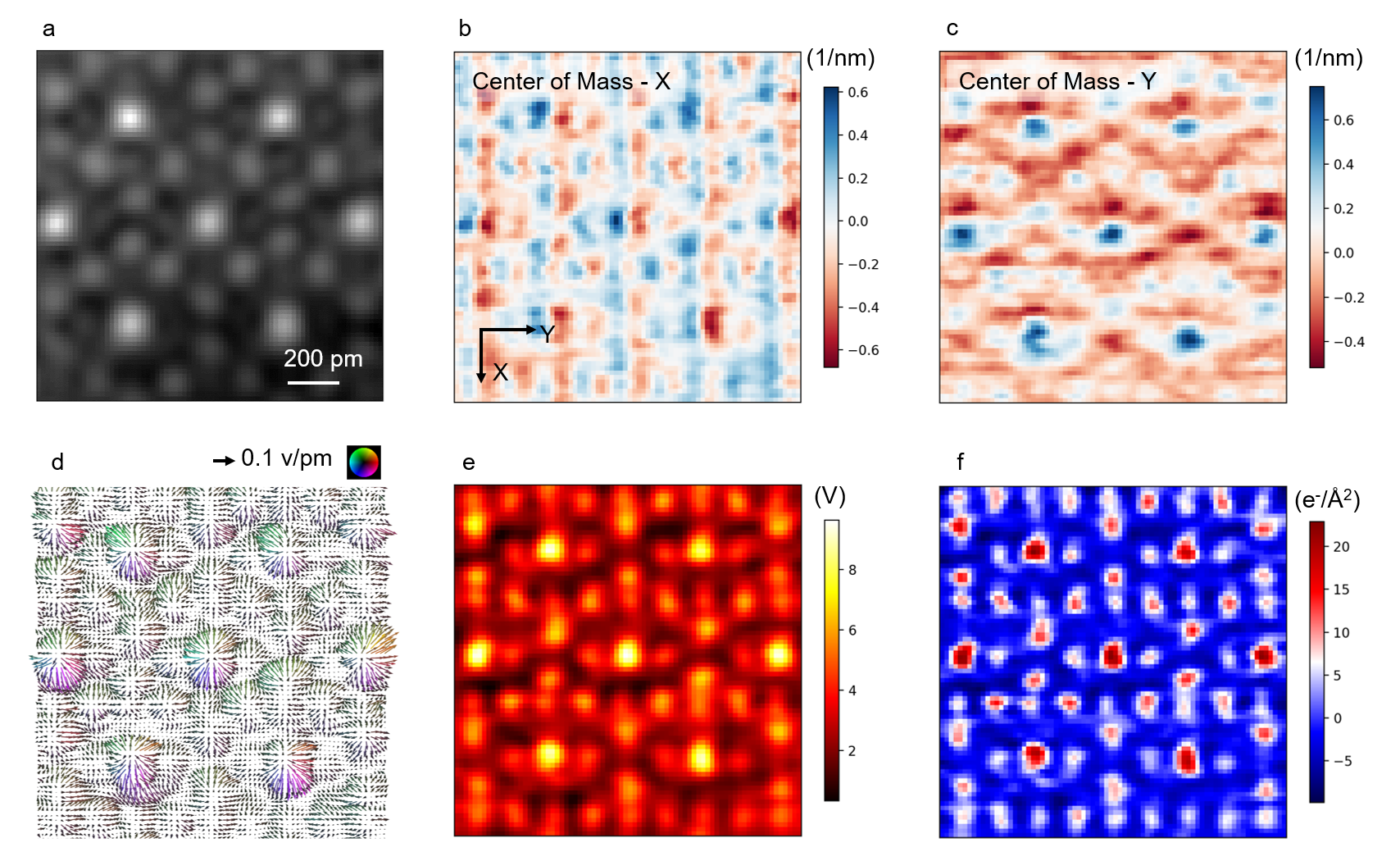}
    \caption{\textbf{Experimental DPC-4DSTEM reconstruction for {\magnetite} oriented in the \hkl[110] direction.} 
    \textbf{a} Reconstructed virtual dark-field image. 
    Change of the center of mass of the transmitted beam in \textbf{b} X and \textbf{c} Y directions. 
    \textbf{d} Electric field vector map. 
    \textbf{e} Projected electrostatic potential map. 
    \textbf{f} Charge-density map. 
    The scanning step size used in this experiment is 13\,pm.}
    \label{fig:DPC_M13}
\end{figure}

\begin{figure}[!htb]
    \centering
    \includegraphics[trim = 0mm 0mm 0mm 0mm, clip, width=1\linewidth]{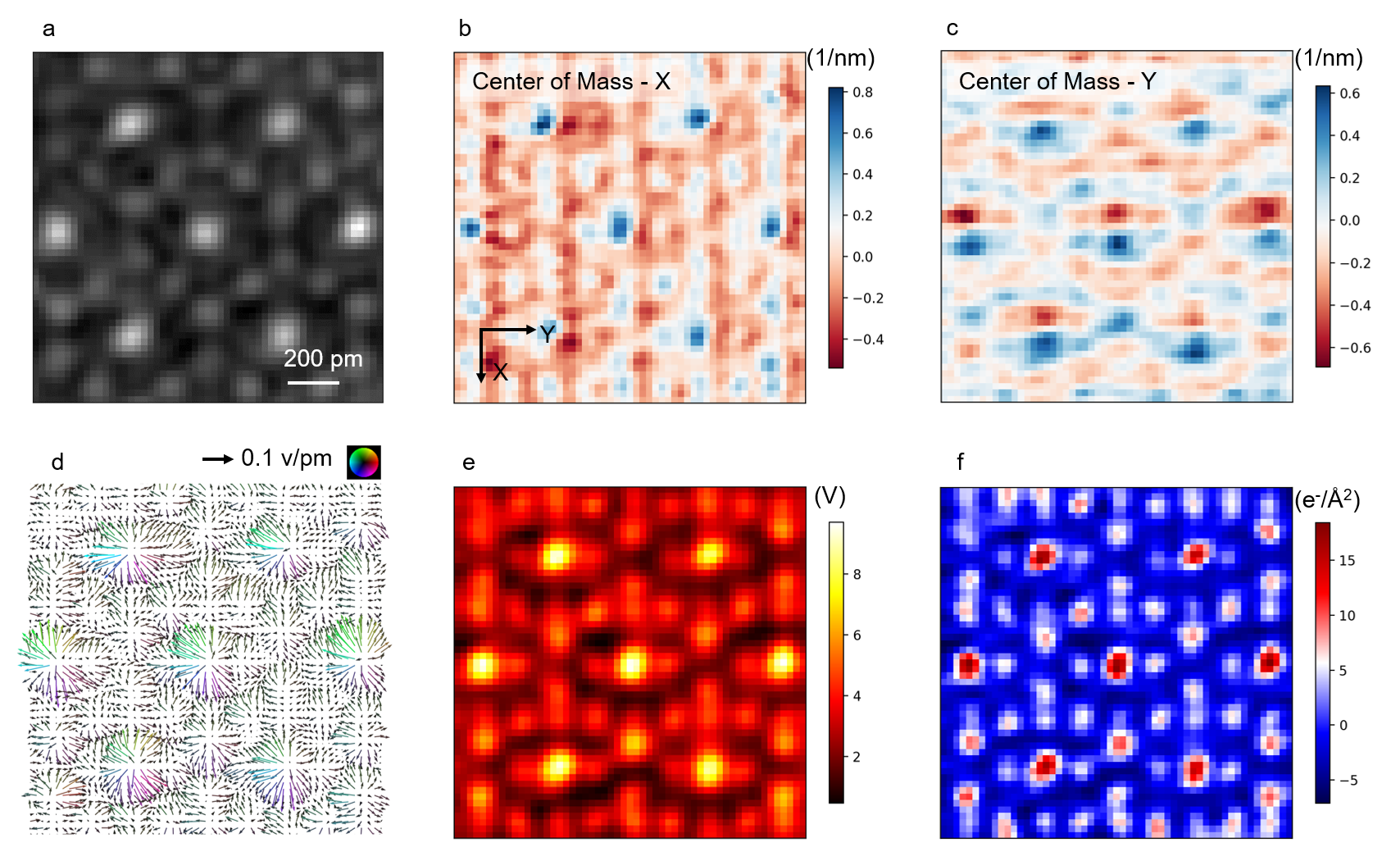}
    \caption{\textbf{Experimental DPC-4DSTEM reconstruction for {\magnetite} oriented in the [110] direction.} 
    \textbf{a} Reconstructed virtual dark-field image. 
    Change of the center of mass of the transmitted beam in \textbf{b} X and \textbf{c} Y directions. 
    \textbf{d} Electric field vector map. 
    \textbf{d} Projected electrostatic potential map.
    \textbf{f} Charge-density map. 
    The scanning step size used in this experiment is 18\,pm.}
    \label{fig:DPC_M18}
\end{figure}

\begin{figure}[!htb]
    \centering
    \includegraphics[trim = 0mm 0mm 0mm 0mm, clip, width=1\linewidth]{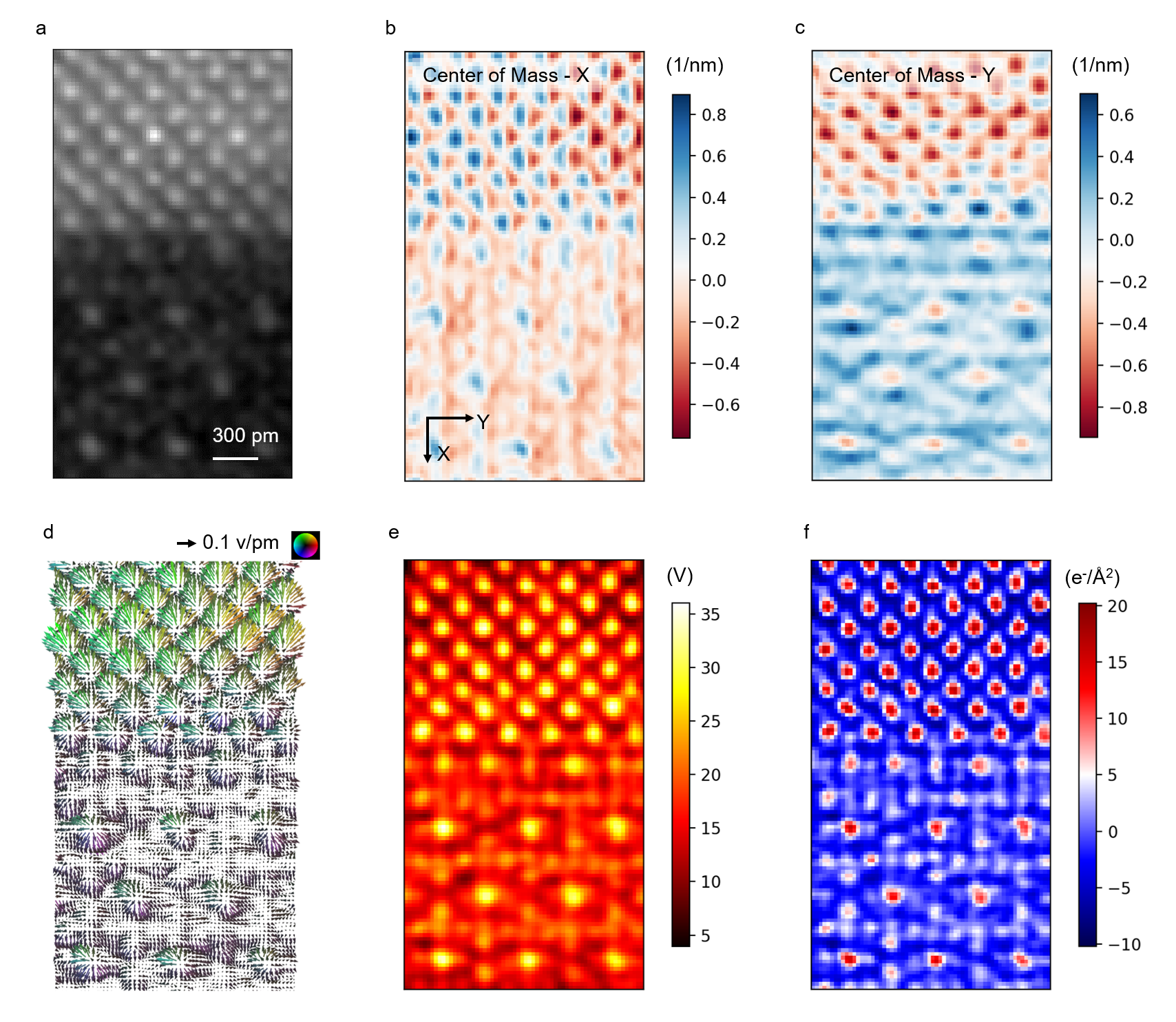}
    \caption{\textbf{Experimental DPC-4DSTEM reconstruction for the Fe/{\magnetite} interface with the {\magnetite} oriented in the \hkl[110] direction.} 
    \textbf{a} Reconstructed virtual dark-field image. 
    Change of the center of mass of the transmitted beam in \textbf{b} X and \textbf{c} Y directions. 
    \textbf{d} Electric field vector map. 
    \textbf{e} Projected electrostatic potential map. 
    \textbf{f} Charge-density map. 
    The scanning step size used in this experiment is 18\,pm.}
    \label{fig:DPC_IM}
\end{figure}

\begin{figure}[!htb]
    \centering
    \includegraphics[trim = 0mm 0mm 0mm 0mm, clip, width=1\linewidth]{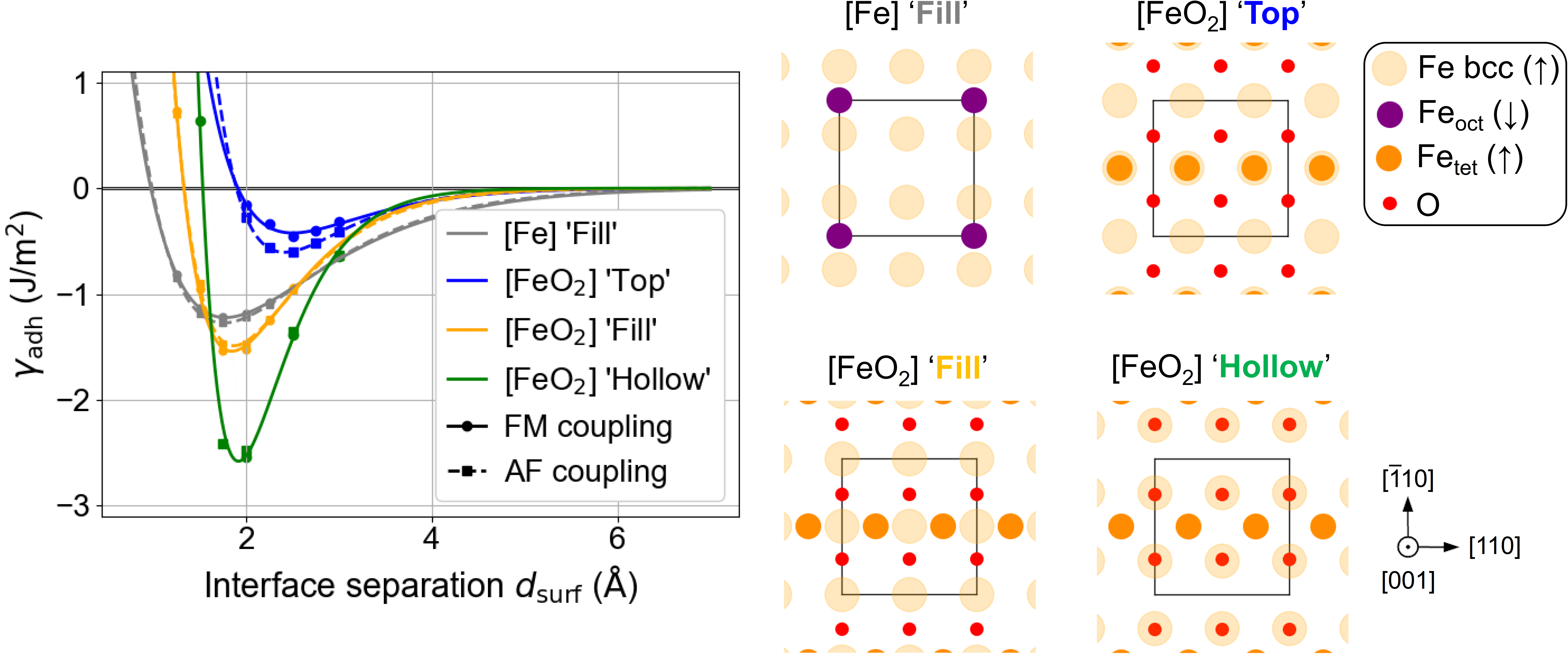}
    \caption{\textbf{Stable sites for the Fe/{\magnetite} interface} Adhesion energy $\gamma_{\mathrm{adh}}$ of the Fe\hkl(001)-{\magnetite}\hkl(001) interface as a function of the interface separation distance $d_{\mathrm{surf}}$ for different relative positions of the two layers. The corresponding sites are shown on the right, viewed along the \hkl[001] direction of the two structures.}
    \label{fig:fig_sites_Fe_Fe3O4}
\end{figure}

\begin{figure}[!htb]
    \centering
    \includegraphics[trim = 1mm 0mm 0mm 0mm, clip, width=1\linewidth]{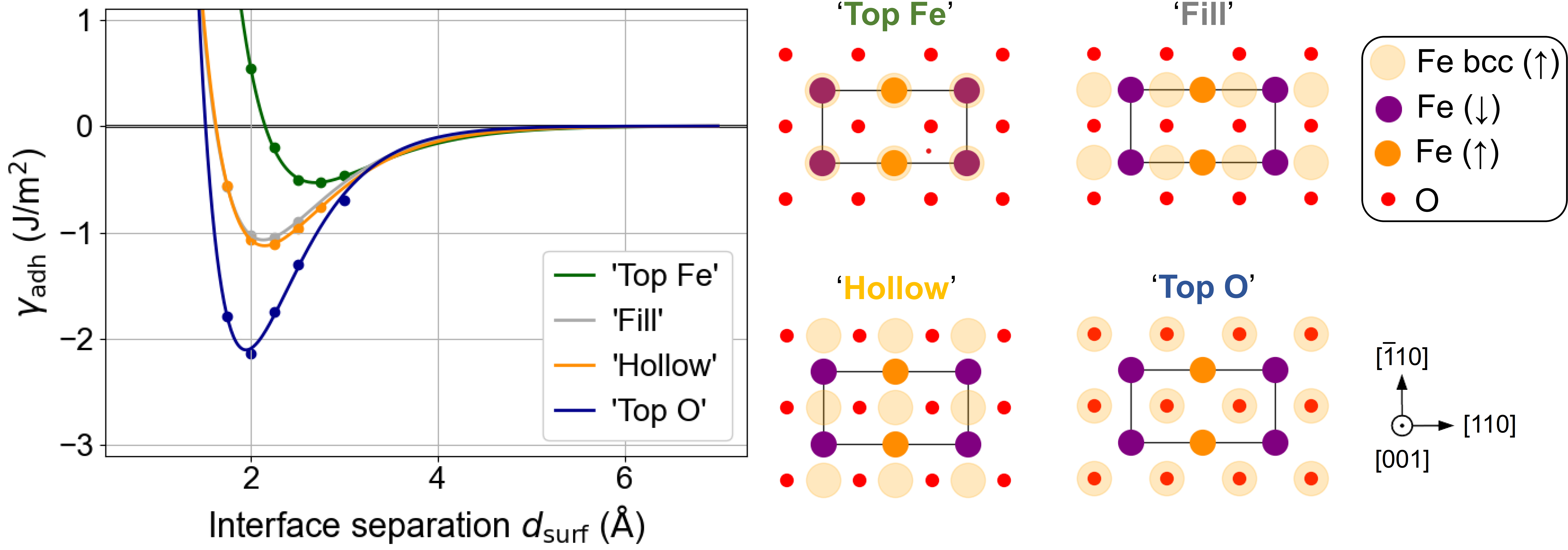}
    \caption{\textbf{Stable sites for the Fe/{\wustiteSlab} interface} Adhesion energy $\gamma_{\mathrm{adh}}$ of the Fe\hkl(001)-{\wustiteSlab}\hkl(001) interface as a function of the interface separation distance $d_{\mathrm{surf}}$ for different relative positions of the two layers. The corresponding sites are shown on the right, viewed along the \hkl[001] direction of the two structures.}
    \label{fig:fig_sites_Fe_FeO}
\end{figure}

\begin{figure}[!htb]
    \centering
    \includegraphics[trim = 0mm 0mm 0mm 0mm, clip, width=1\linewidth]{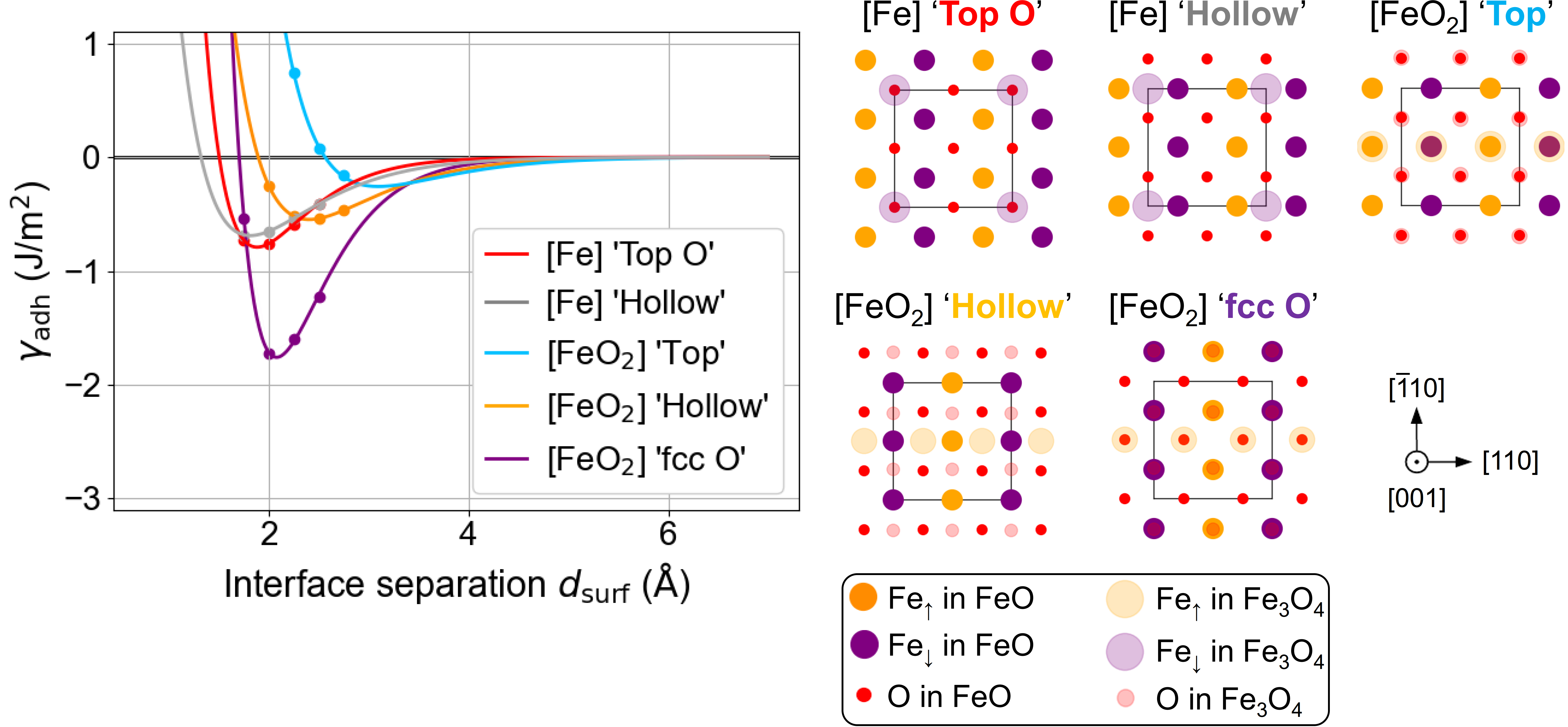}
    \caption{\textbf{Stable sites for the {\wustiteSlab}-{\magnetite} interface} Adhesion energy $\gamma_{\mathrm{adh}}$ of the {\wustiteSlab}\hkl(001)-{\magnetite}\hkl(001) interface as a function of the interface separation distance $d_{\mathrm{surf}}$ for different relative positions of the two layers. The corresponding sites are shown on the right, viewed along the \hkl[001] direction of the two structures.}
    \label{fig:fig_sites_FeO_Fe3O4}
\end{figure}

\begin{figure}[!htb]
    \centering
    \includegraphics[trim = 0mm 0mm 0mm 0mm, clip, width=0.85\linewidth]{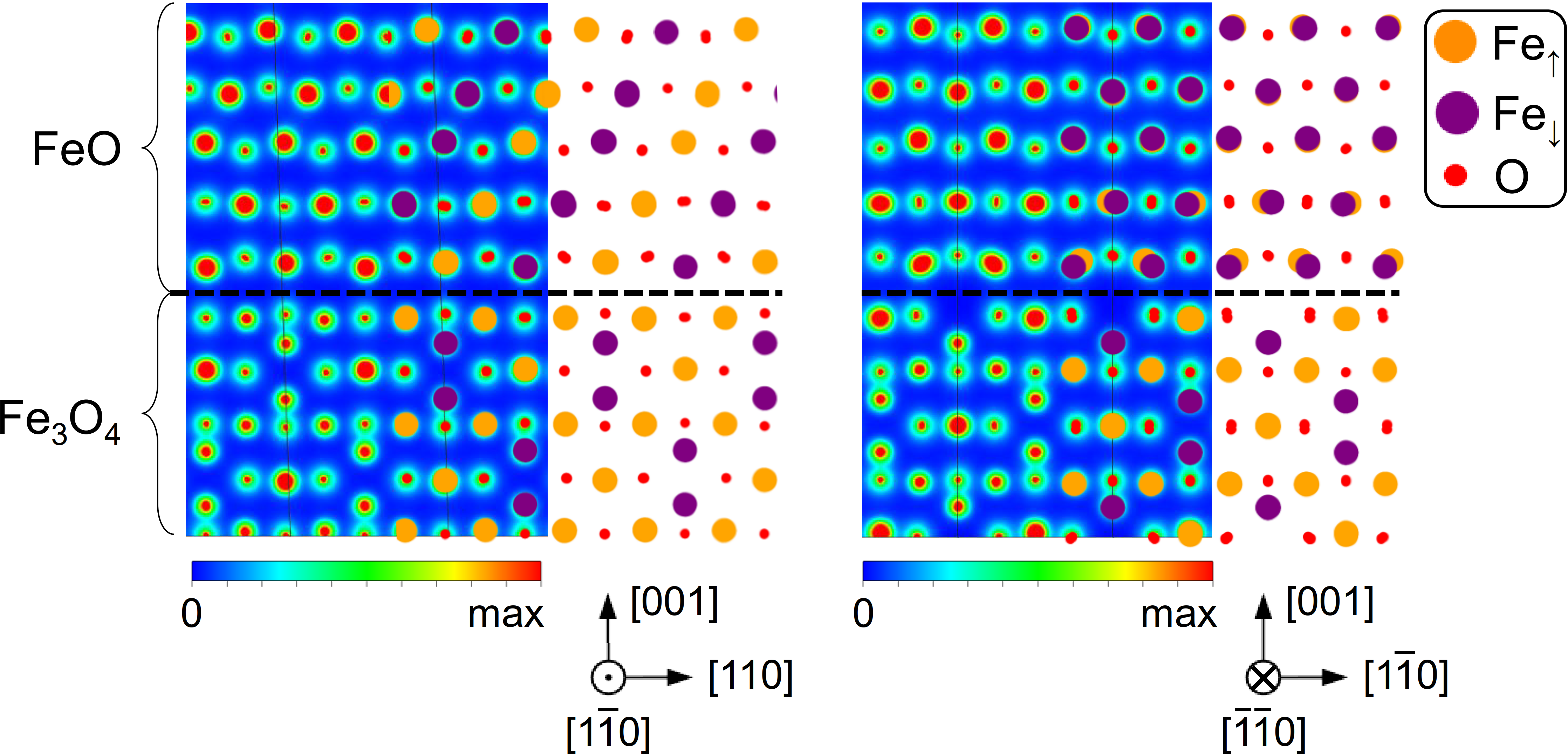}
    \caption{\textbf{The relaxed {\wustiteSlab}-{\magnetite} interface} Charge density and relaxed atomic structure of the {\wustiteSlab}\hkl(001)-{\magnetite}\hkl(001) interface obtained using DFT calculations. With respect to \fig{fig:fig_sites_FeO_Fe3O4}{}, the configuration of the interface corresponds to the [FeO$_2$] ``fcc O'' site, where the two fcc O sub-lattices of {\wustiteSlab} and {\magnetite} match along their respective \hkl[001] direction.}
    \label{fig:fig_charge_FeO_Fe3O4}
\end{figure}

\begin{figure}[!htb]
    \centering
    \includegraphics[trim = 0mm 0mm 0mm 0mm, clip, width=1\linewidth]{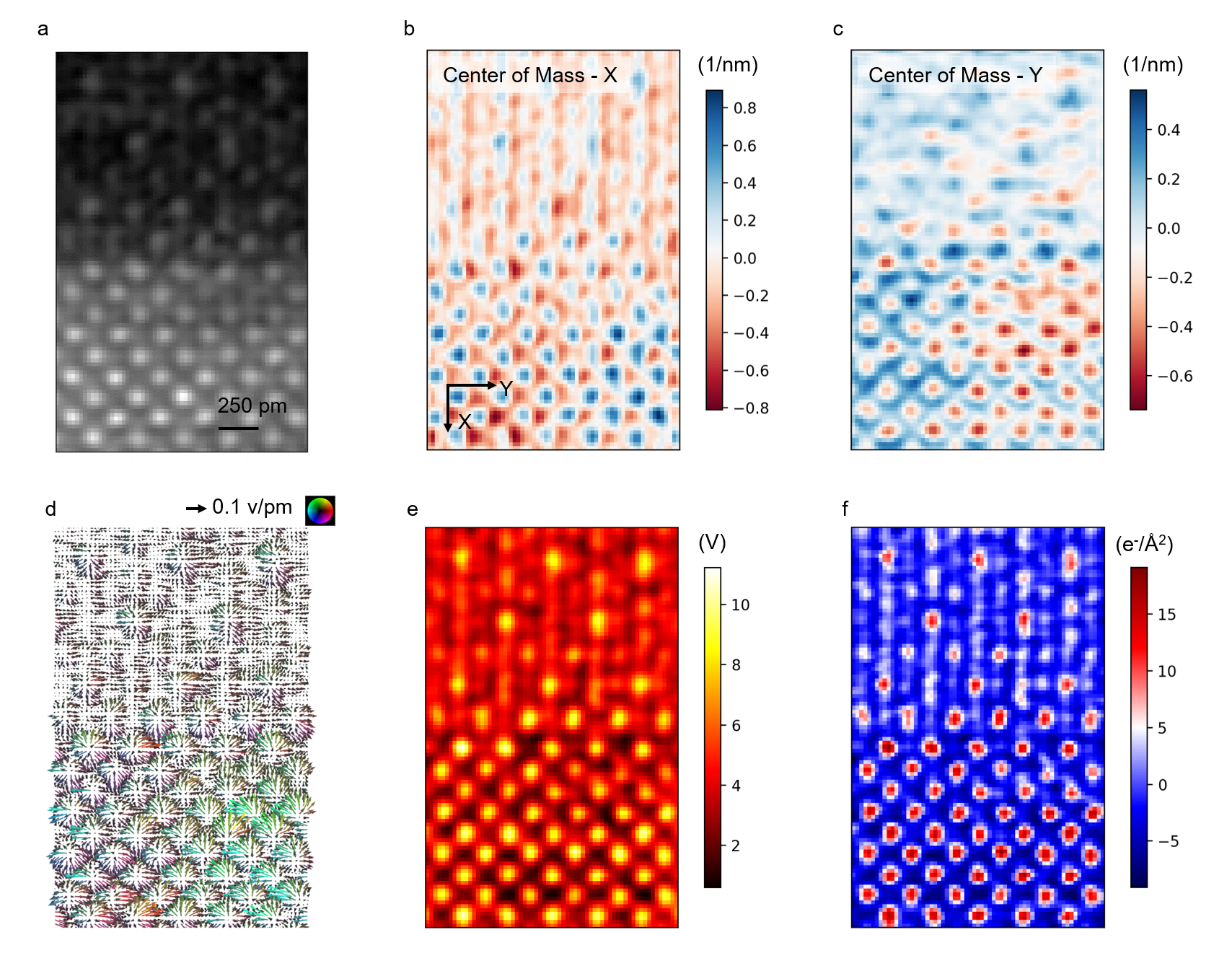}
    \caption{\textbf{Experimental DPC-4DSTEM reconstruction for the {\magnetite}-Fe interface with the {\magnetite} oriented in the \hkl[110] direction.} 
    \textbf{a} Reconstructed virtual dark-field image. 
    Change of the center of mass of the transmitted beam in \textbf{b} X and \textbf{c} Y directions. 
    \textbf{d} Electric field vector map. 
    \textbf{e} Projected electrostatic potential map. 
    \textbf{f} Charge-density map. 
    The scanning step size used in this experiment is 18\,pm.}
    \label{fig:DPC_MI}
\end{figure}

\begin{figure}[!htb]
    \centering
    \includegraphics[trim = 0mm 0mm 3mm 0mm, clip, width=0.85\linewidth]{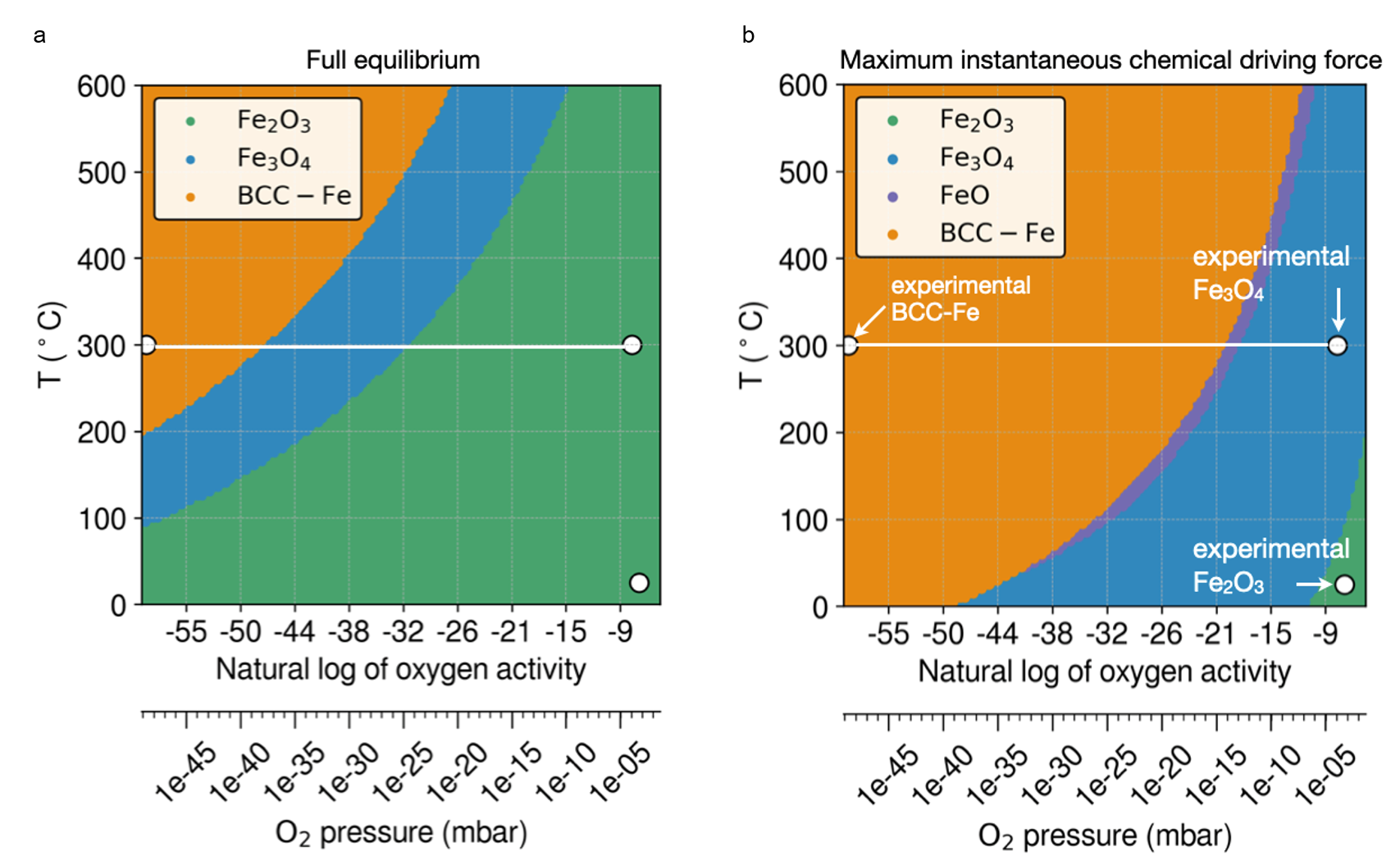}
    \caption{\textbf{Bulk thermodynamics in thin film interface fabrication.}
    \textbf{a} Equilibrium phase diagram as a function of O activity and temperature;
    \textbf{b} The maximum instantaneous chemical driving force diagram illustrates the phases with the maximum instantaneous chemical driving force as a function of O activity and temperature.
    Three distinct external O activities lead to three possible syntheses of phases.
    A detailed presentation of the maximum instantaneous chemical driving force mapping is shown in \fig{fig:Supp_TC}{}.
    }
    \label{fig:Thermo}
\end{figure}

\begin{figure}[!htb]
    \centering
    \includegraphics[trim = 0mm 0mm 0mm 0mm, clip, width=1\linewidth]{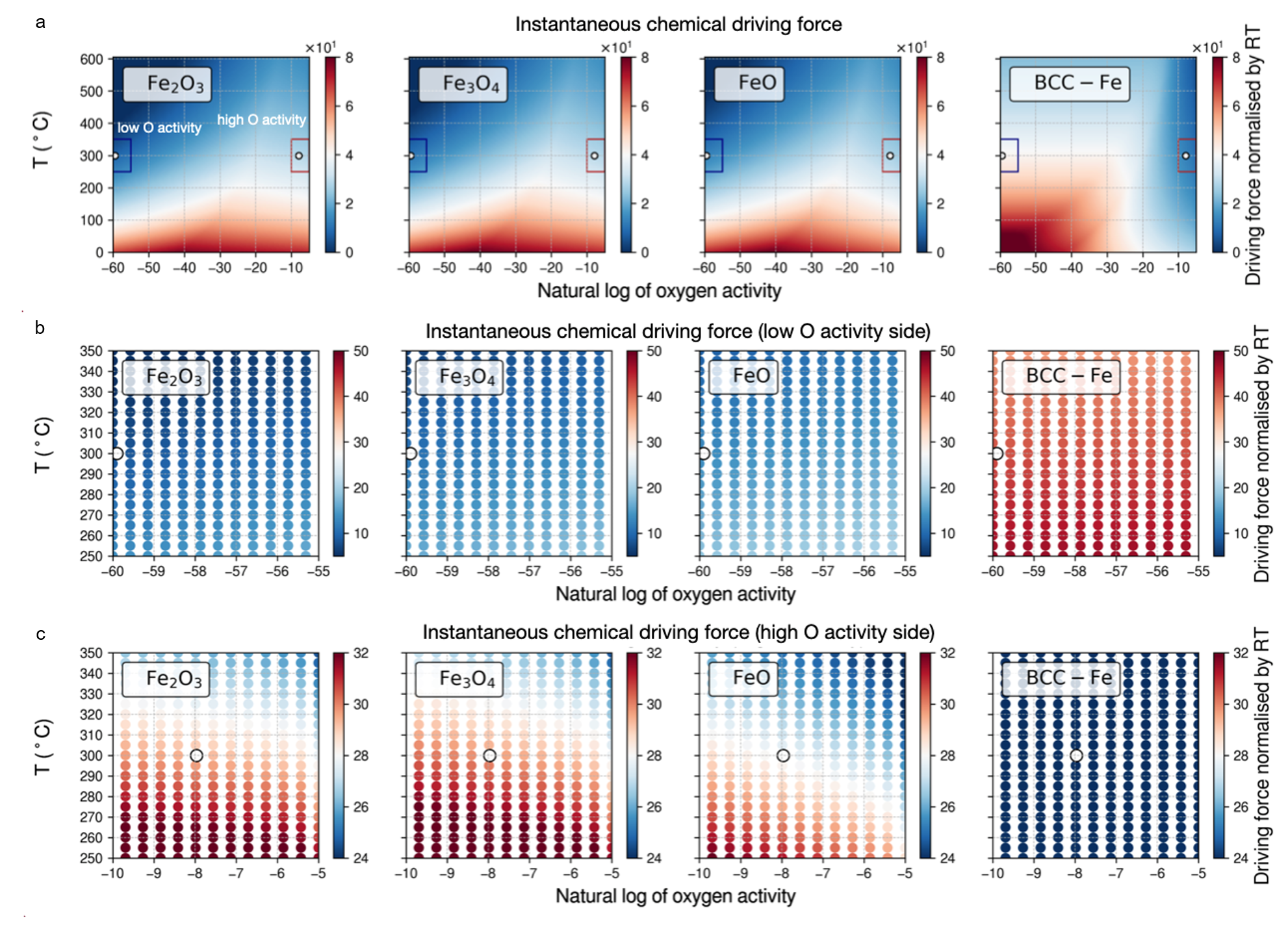}
    \caption{
    \textbf{The maximum instantaneous chemical driving force as a function of temperature and O activity.} 
    \textbf{a} Mapping of the maximum instantaneous chemical driving force for the deposition of solid phases from the gas phase.
    \textbf{b \& c} Enlarged view of the maximum instantaneous chemical driving force landscape in the region of low O activity and high O activity regimens respectively, as highlighted in \textbf{a}.
    }
    \label{fig:Supp_TC}
\end{figure}

\begin{figure}[!htb]
    \centering
    \includegraphics[trim = 0mm 0mm 0mm 0mm, clip, width=1\linewidth]{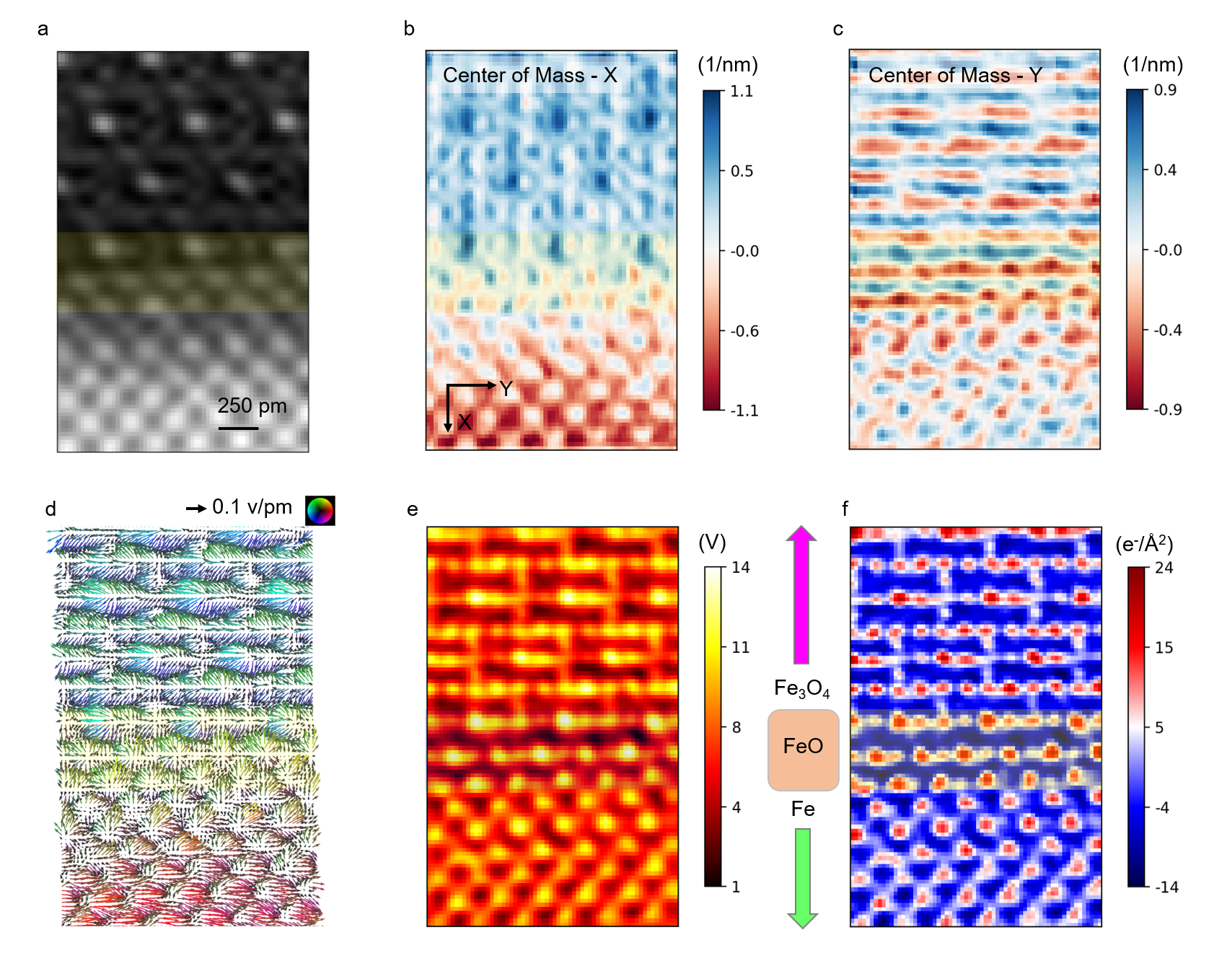}
    \caption{\textbf{Experimental DPC-4DSTEM reconstruction for a three-layer thick {\wustiteSlab} slab at the Fe/{\magnetite} interface, with \magnetite oriented in the \hkl[110] direction.} 
    \textbf{a} Reconstructed virtual dark-field image. 
    Change of the center of mass of the transmitted beam in \textbf{b} X and \textbf{c} Y directions. 
    \textbf{d} Electric field vector map. 
    \textbf{e} Projected electrostatic potential map. 
    \textbf{f} Charge-density map. 
    The scanning step size used in this experiment is 18\,pm.
    The specimen is the same as shown in \fig{fig:in-situ}{}, which underwent in-situ heating in TEM at \SI{300}{{\degree}C} for \SI{2}{\hour}.}
    \label{fig:DPC_MI_3layer}
\end{figure}

\begin{figure}[!htb]
    \centering
    \includegraphics[trim = 0mm 0mm 0mm 0mm, clip, width=1\linewidth]{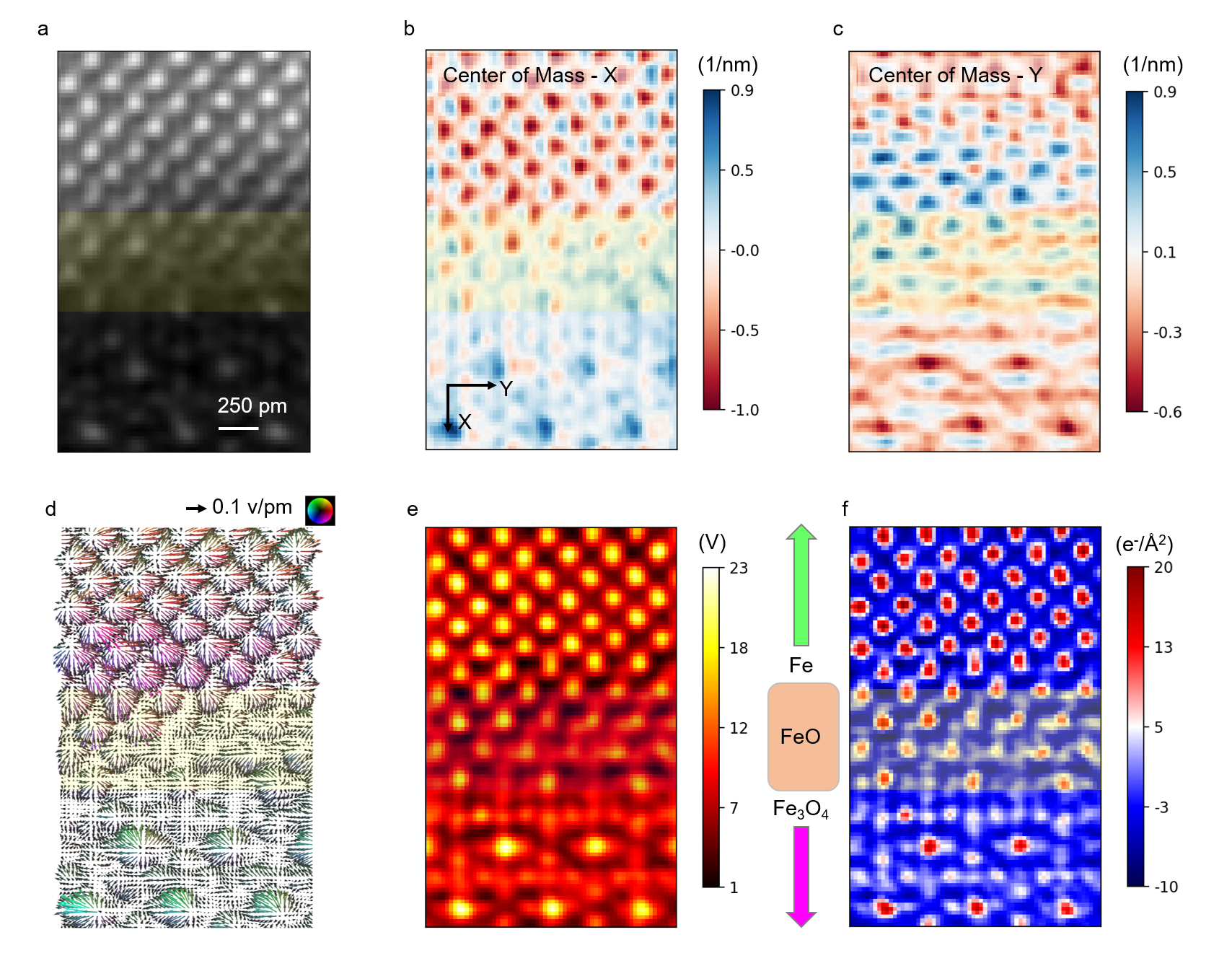}
    \caption{\textbf{Experimental DPC-4DSTEM reconstruction for a four-layer thick {\wustiteSlab} slab at the Fe/{\magnetite} interface, with \magnetite oriented in the \hkl[110] direction.} 
    \textbf{a} Reconstructed virtual dark-field image. 
    Change of the center of mass of the transmitted beam in \textbf{b} X and \textbf{c} Y directions. 
    \textbf{d} Electric field vector map. 
    \textbf{e} Projected electrostatic potential map. 
    \textbf{f} Charge-density map. 
    The scanning step size used in this experiment is 18\,pm.
    }
    \label{fig:DPC_IM_4layer}
\end{figure}

\begin{figure}[!htb]
    \centering
    \includegraphics[trim = 0mm 0mm 0mm 0mm, clip, width=1\linewidth]{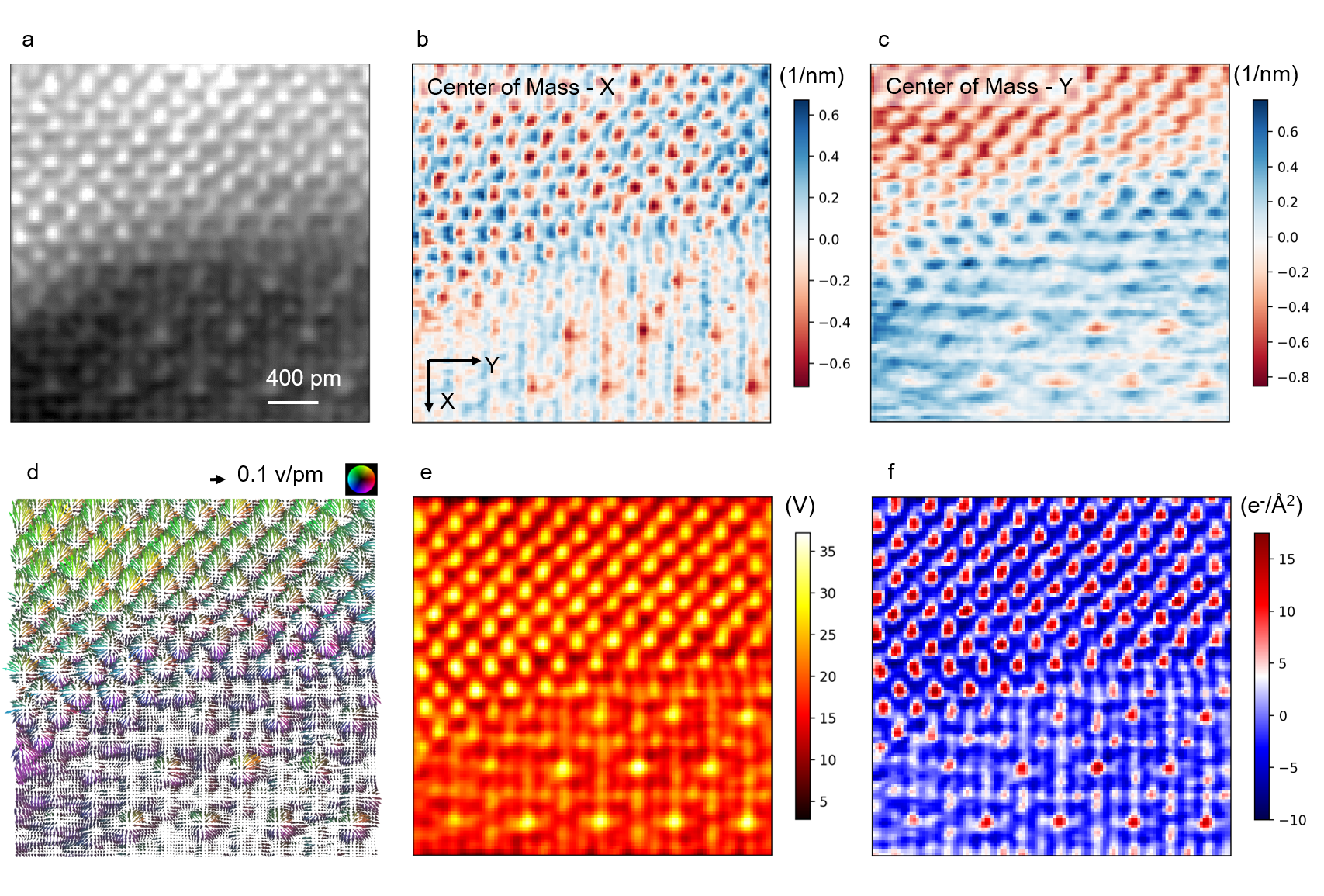}
    \caption{\textbf{Experimental DPC-4DSTEM reconstruction for the Fe/{\magnetite} interface with a step, {\magnetite} oriented in the \hkl[110] direction.} 
    \textbf{a} Reconstructed virtual dark-field image. 
    Change of the center of mass of the transmitted beam in \textbf{b} X and \textbf{c} Y directions. 
    \textbf{d} Electric field vector map. 
    \textbf{e} Projected electrostatic potential map. 
    \textbf{f} Charge-density map. 
    The scanning step size used in this experiment is 18\,pm.}
    \label{fig:DPC_Step}
\end{figure}

\begin{figure}[!htb]
    \centering
    \includegraphics[trim = 0mm 0mm 0mm 0mm, clip, width=1\linewidth]{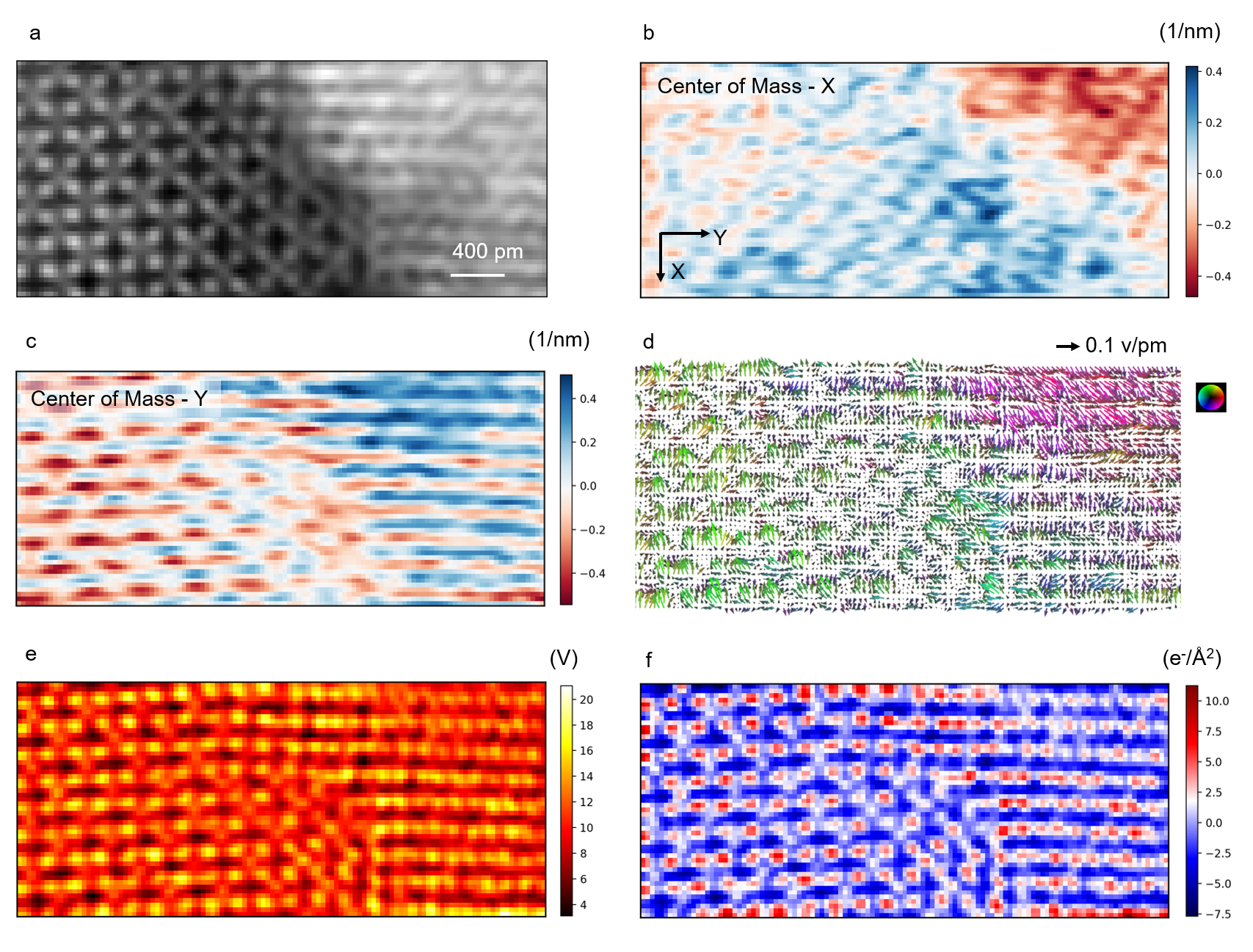}
    \caption{\textbf{Experimental DPC-4DSTEM reconstruction of the semi-coherent interface between Fe and {\magnetite}, with {\magnetite} oriented in the \hkl[100] direction.} 
    \textbf{a} Reconstructed virtual dark-field image. 
    Change of the center of mass of the transmitted beam in \textbf{b} X and \textbf{c} Y directions. 
    \textbf{d} Electric field vector map. 
    \textbf{e} Projected electrostatic potential map. 
    \textbf{f} Charge-density map. 
    The scanning step size used in this experiment is 25\,pm.}
    \label{fig:DPC_Semico}
\end{figure}

\begin{figure}[!htb]
    \centering
    \includegraphics[trim = 0mm 0mm 0mm 0mm, clip, width=1\linewidth]{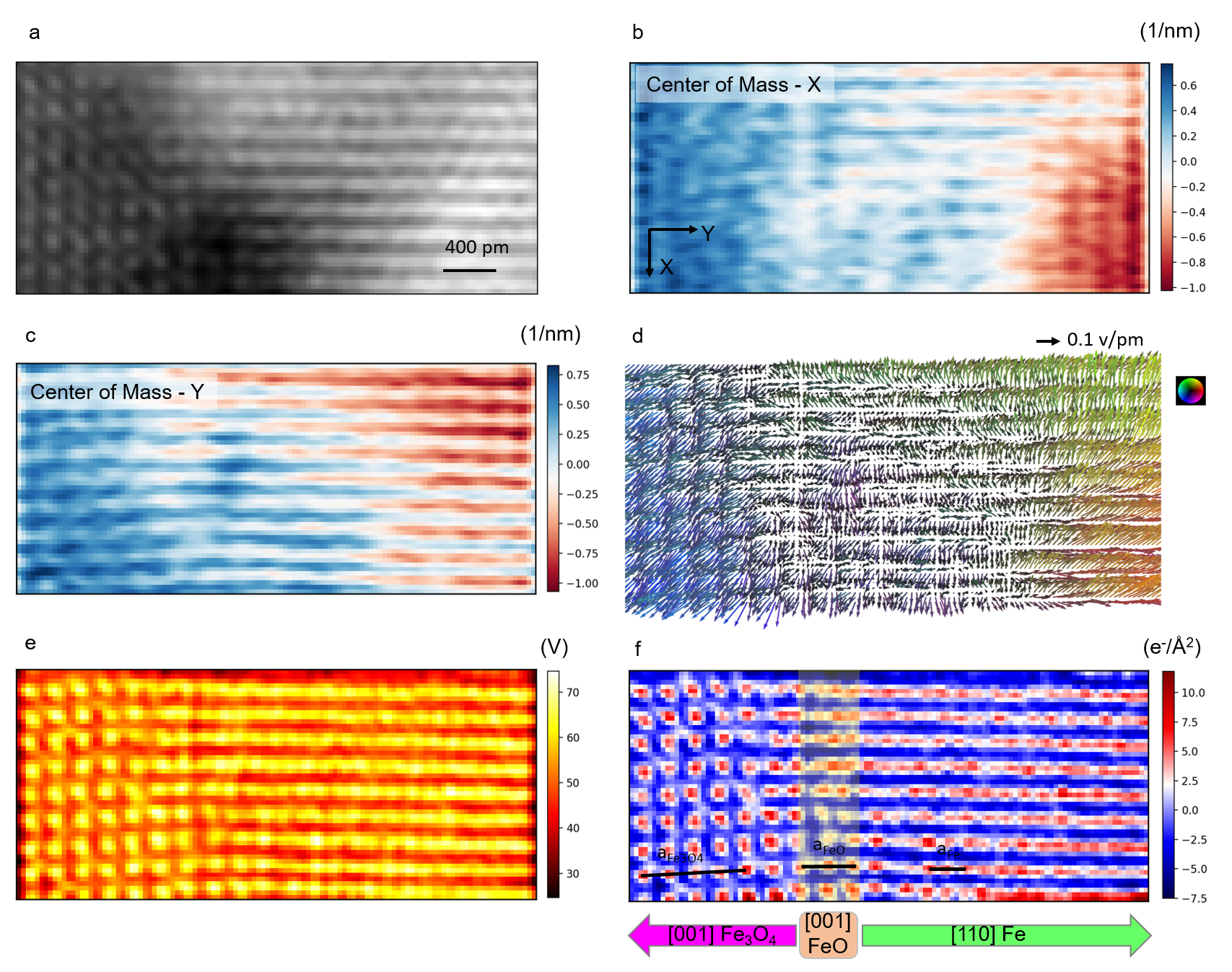}
    \caption{\textbf{Experimental DPC-4DSTEM reconstruction of the coherent interface between Fe and {\magnetite}, with {\magnetite} oriented in the \hkl[100] direction.} 
    \textbf{a} Reconstructed virtual dark-field image. 
    Change of the center of mass of the transmitted beam in \textbf{b} X and \textbf{c} Y directions. 
    \textbf{d} Electric field vector map. 
    \textbf{e} Projected electrostatic potential map. 
    \textbf{f} Charge-density map. 
    The scanning step size used in this experiment is 25\,pm.
    \RV{Approximate lattice parameters ($a_{Fe3O4}$, $a_{FeO}$ and $a_{Fe}$) are highlighted as black lines between layers of atoms.}
    }
    \label{fig:DPC_Coherent}
\end{figure}

\begin{figure}[!htb]
    \centering
    \includegraphics[trim = 0mm 0mm 0mm 0mm, clip, width=1\linewidth]{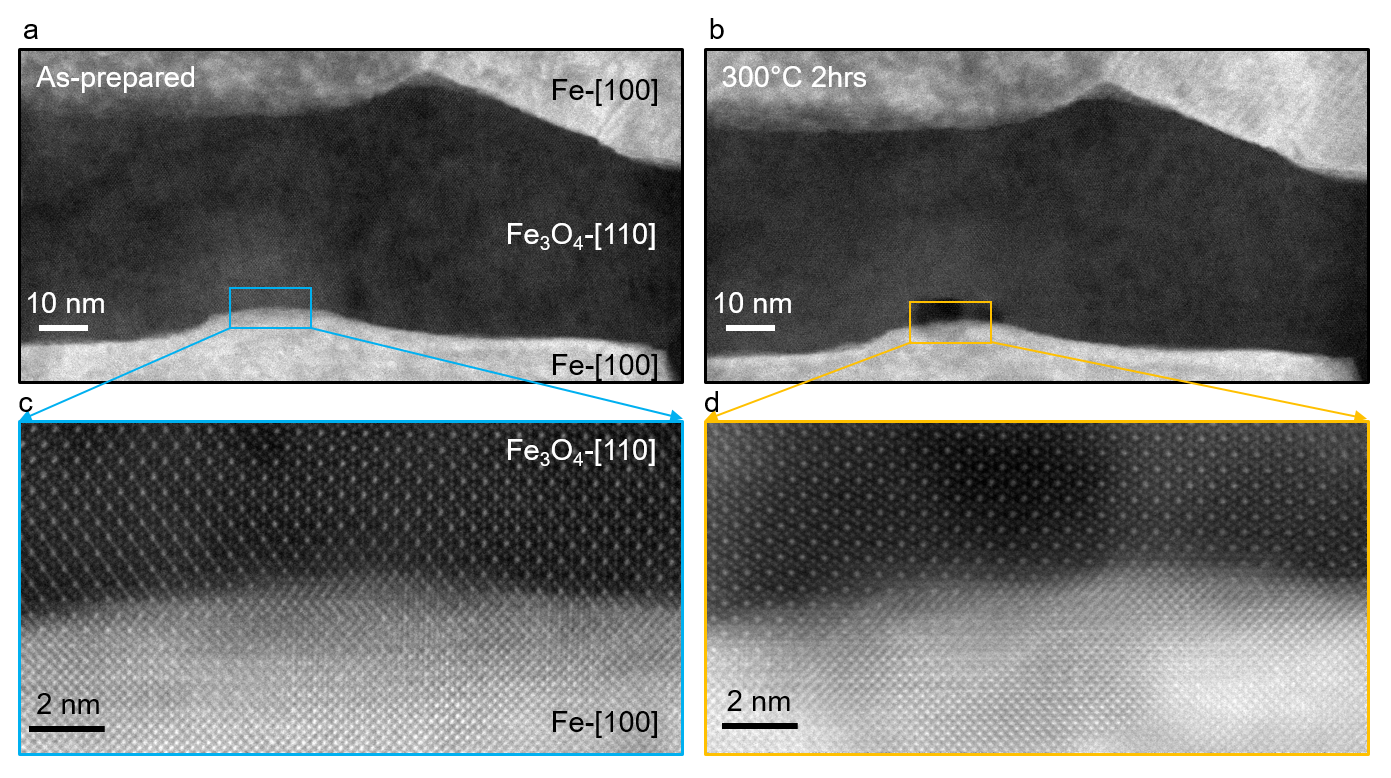}
    \caption{\textbf{Comparison of the interface structure between Fe and {\magnetite} before and after in-situ heating experiments.} 
    \textbf{a} Overview HAADF imaging of the Fe-{\magnetite}-Fe structure in the as-prepared condition. 
    \textbf{b} The same structure after in-situ heating in TEM at \SI{300}{{\degree}C} for \SI{2}{\hour}, with a heating and cooling rate of \SI{5}{{\degree}C/\second}.
    Both images were taken at room temperature. 
    \textbf{c \& d} Magnified HAADF images that show the atomic structure of the interface between \magnetite-\hkl[110] and Fe-\hkl[100].}
    \label{fig:in-situ}
\end{figure}

\begin{figure}[!htb]
    \centering
    \includegraphics[trim = 0mm 0mm 0mm 0mm, clip, width=1\linewidth]{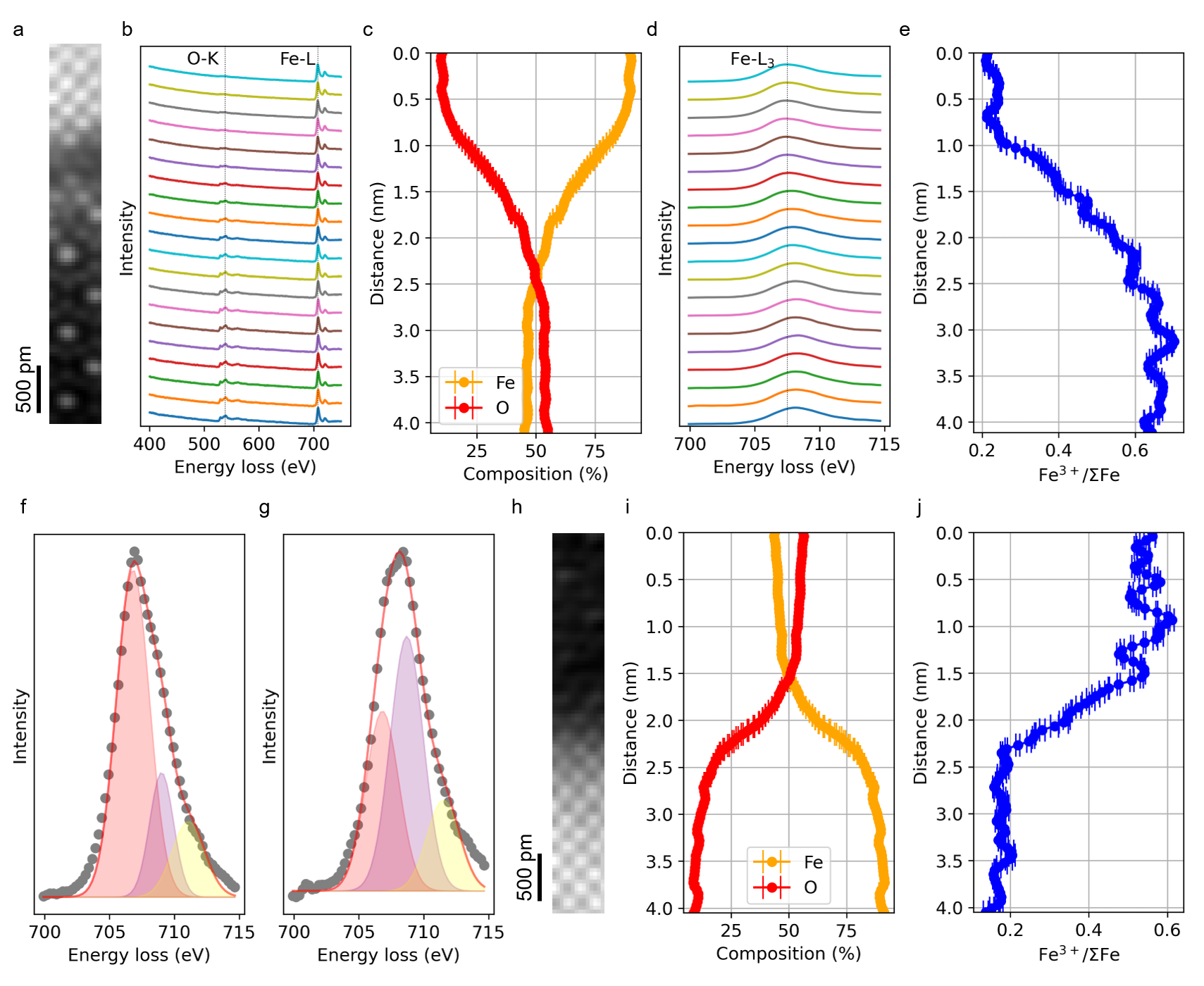}
    \caption{\textbf{Local chemistry and charge state analysis for complexions at the Fe-{\magnetite} interface.} 
    \textbf{a} Dark-field image serves for electron energy loss spectroscopy (EELS) analysis of the Fe-Fe3O4 interface. 
    \textbf{b} Selected energy loss spectra for scanning regions across the Fe-{\magnetite} interface correspond to regions in the dark-field image from a.
    The spectra highlight the O-K edge and Fe-L edges.
    \textbf{c} Quantified local composition using EELS for regions depicted in a, with distances from \SI{0}{nm}-\SI{4.2}{nm} matching the dark-field image from top to bottom.
    \textbf{d} Magnified regions of the spectra from b show the Fe-$L_3$ peak, revealing a shift to a higher energy state from the Fe region (upper part) to the {\magnetite} (lower part) region. 
    \textbf{e} Quantified the charge state using EELS for the regions shown in a. 
    \textbf{f \& g} display selected spectra for peak decomposition in the Fe region and the {\magnetite} region, respectively, using three Gaussian peaks (pink, purple, and yellow) for decomposition. 
    The ratio of $Fe^{3+}$/$\Sigma$Fe inversely relates to the integrated area of the pink peak \cite{van2002EELS}. 
    \textbf{h-j} present the dark-field image for the EELS analysis and quantification of local composition and charge state for the scanning regions across the {\magnetite}-Fe interface.
    }
    \label{fig:EELS}
\end{figure}

\begin{figure}[!htb]
    \centering
    \includegraphics[trim = 0mm 0mm 0mm 0mm, clip, width=1\linewidth]{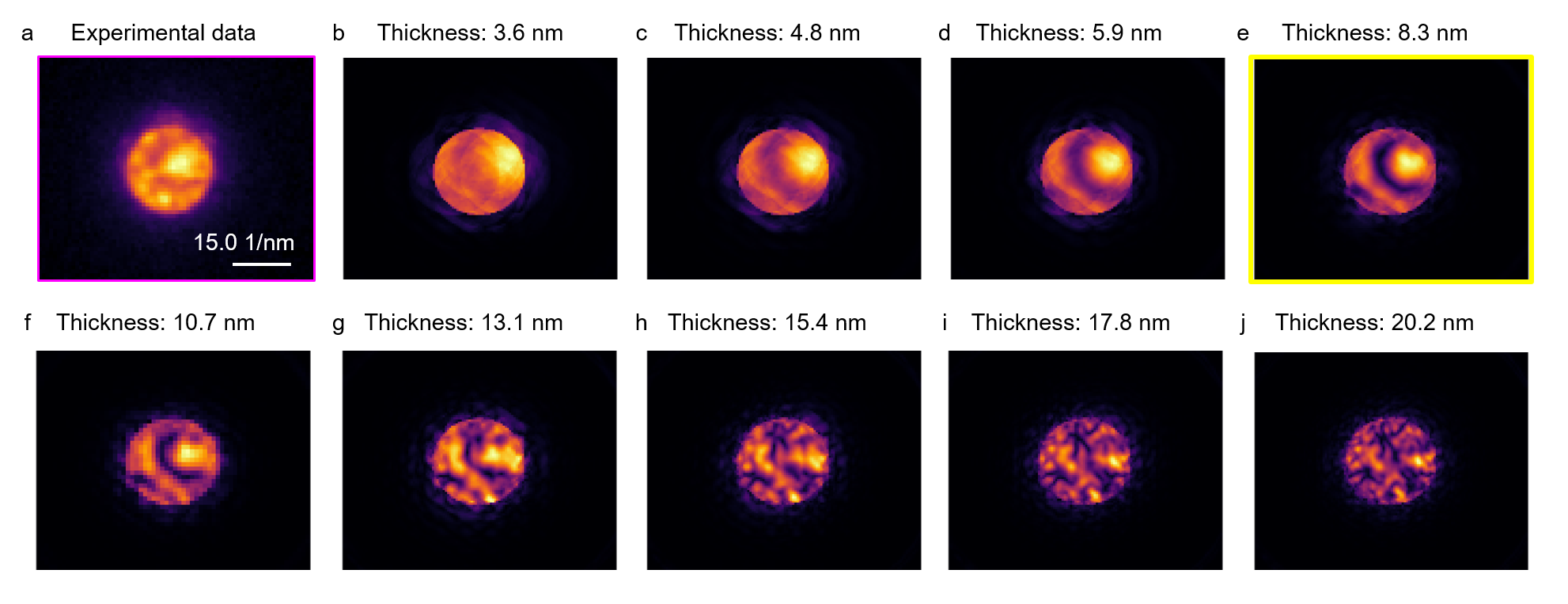}
    \caption{\textbf{Measurement of sample thickness by convergent beam electron diffraction (CBED).} 
    \textbf{a} displays an example of an experimental CBED pattern. 
    \textbf{b-j} include similar simulated CBED patterns for a range of thicknesses from 3.6\,nm to 20.2\,nm.}
    \label{fig:DP_Thickness}
\end{figure}

\begin{figure}[!htb]
    \centering
    \includegraphics[trim = 0mm 0mm 0mm 0mm, clip, width=1\linewidth]{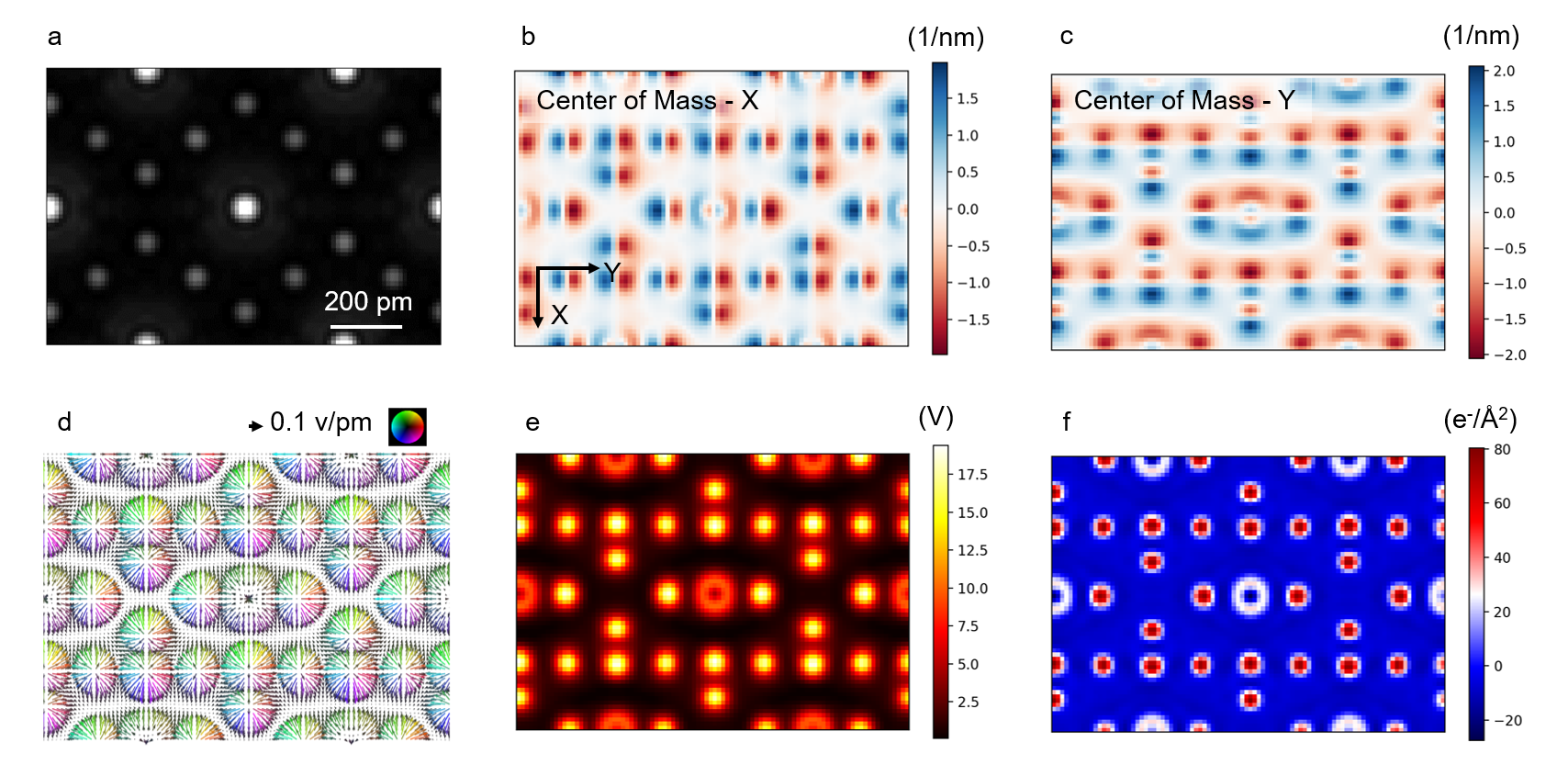}
    \caption{\textbf{Simulated DPC-4DSTEM reconstruction for {\magnetite} oriented in the [110] direction.} 
    \textbf{a} Reconstructed virtual dark-field image. 
    Change of the center of mass of the transmitted beam in \textbf{b} X and \textbf{c} Y directions. 
    \textbf{d} Electric field vector map. 
    \textbf{e} Projected electrostatic potential map. 
    \textbf{f} Charge-density map. 
    The scanning step size used in this experiment is 12\,pm.}
    \label{fig:mSTEM_M}
\end{figure}

\begin{figure}[!htb]
    \centering
    \includegraphics[trim = 0mm 0mm 0mm 0mm, clip, width=1\linewidth]{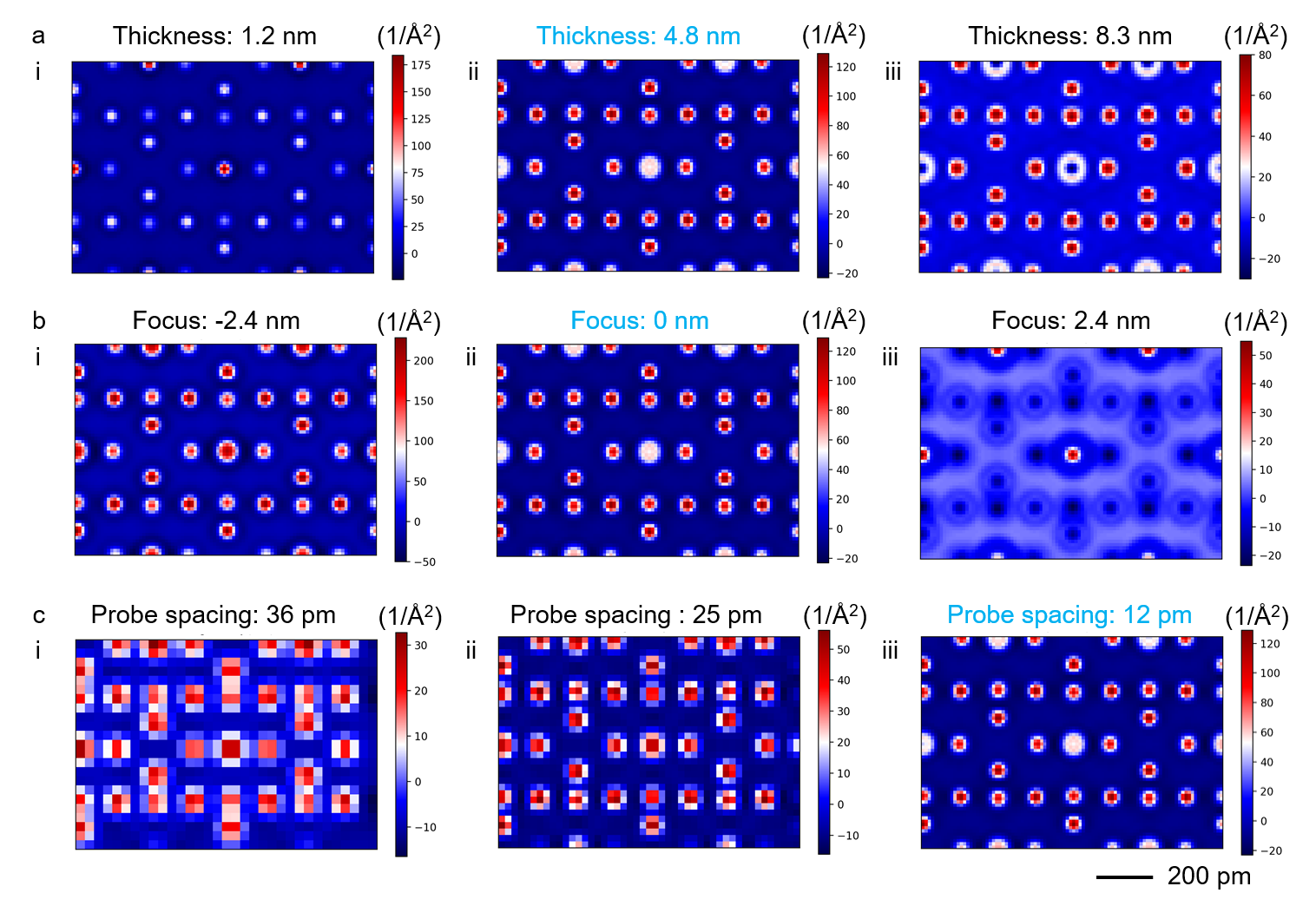}
    \caption{\textbf{The effect of scanning parameters on the reconstructed charge-density maps}:
    \textbf{a-i-iii}  show the effect of sample thickness, ranging from 1.2\,nm to 8.3\,nm. 
    \textbf{b-i-iii} illustrate the impact of defocus, ranging from -2.4\,nm to 2.4\,nm. 
    \textbf{c-i-iii} demonstrate the influence of probe spacing, ranging from 36\,pm to 12\,pm. 
    All charge-density maps were reconstructed from simulated DPC-4DSTEM data of the {\magnetite} crystal oriented in the \hkl(110) direction.}
    \label{fig:mSTEM_Para}
\end{figure}

\end{appendices}

\clearpage


\end{document}